\documentclass{article}
\usepackage{graphicx}
\usepackage{amssymb,amsmath}
\usepackage{amsfonts}
\begin{document}

\title{High $T_C$ Superconductivity: Doping Dependent Theory Confirmed by Experiment}
\author{Stephen B. Haley\thanks{profisbh@gmail.com}\quad and \quad
 Herman J. Fink\thanks{hjfink@ucdavis.edu}\\
Department of Electrical and Computer Engineering\\
University of California, Davis, CA 95616, USA\\}
\maketitle
\date{
\begin{abstract}
\normalsize
A Hamiltonian $H(\Gamma)$ applicable to cuprate HTS, with a doping dependent pairing interaction $\Gamma(x) = V(x) + U(x)$, is linked to a Cu3d-O2p state probability model(SPM).  A consequence of doping induced electron hopping, the SPM mandates that plaquettes with net charge and spin form in the CuO plane, establishing an effective spin-singlet exchange interaction $U(x)$.  The $U(x)$ is determined from a set of probability functions that characterize the occupation of the single particle states.   An exact treatment of the average static fluctuation part of $H$ shows that diagonal matrix elements $U_{\bf k k} < 0$ produce very effective pairing, with significant deviation from the mean field approximation, which also depends on a phonon-mediated interaction $V$.  This deviation is primarily responsible for the diverse set of HTS properties.  The SC phase transition boundary $T_C(x)$, the SC gap $\Delta(x)$, and the pseudogap $\Delta_{pg}(x)$ are fundamentally related.  Predictions are in excellent agreement with experiment, and a new class of HTS materials is proposed.  Large static fluctuation results in extreme HTS and quantum criticality.\\

\noindent {\bf PACS numbers}:{74.20.Fg, 74.20.Mn, 74.25.-q, 74.10.+v, 74.62.-c, 74.72.-h\\}

\noindent {\bf Keywords}:{high $T_C$, cuprates, polaron, spin exchange, quantum criticality, state probability}

\end{abstract}}
\setlength{\oddsidemargin}{-0.15in} 
 \setlength{\textwidth}{6.9in} 
  \setlength{\columnwidth}{3.2in}

\twocolumn
\sloppy
\section{INTRODUCTION}
The 1986 discovery of non-elemental high $T_C$ superconductor(HTS) cuprates\cite{muller} introduced a class of antiferromagnetic Mott insulator ceramics that exhibit an exotic array of seemingly disparate superconducting(SC) and normal state(NS) properties.\cite{lee,tsuei}   Despite the plethora of experimental data indicating several microscopic interactions, there is incomplete agreement on which interactions are essential and how they combine to produce HTS.\cite{zaanen}  No current Hamiltonian leads to the diverse HTS properties observed in cuprates.  This paper presents a Hamiltonian, conjoined with a state probability model(SPM), that predicts the observed properties listed below and elucidates a key ingredient for pairing glue.  It is shown that the pseudogap results from doping induced charge-spin fluctuation, and a new class of HTS materials with an anti-pseudogap is suggested.

The disparity between elemental low $T_C < 10$K superconductor(LTS) and HTS properties is remarkable.  The doped cuprates are extreme type-II superconductors, exhibiting strong magnetic field induced quantum fluctuations, with the interpretation that the state of these materials may lie in close proximity to a quantum critical state.\cite{beyer,balakirev}  Denoting $x$ as the hole(electron) doping concentration, nine diverse cuprate HTS properties are: order of magnitude increased a) average energy gap amplitude $\Delta (x, T)$, and b) $T_C(x)$, which is maximum at optimal doping $x_{op}$, can exceed 100K, with $T_C/T_F \sim 10^{-1}-10^{-2}$.  c) large shape ratio $\Delta (x_{op},0)/k_B T_C(x_{op}) \gg 2$, the maximum BCS LTS value,  d) inverted parabolic(dome) shaped phase boundary $T_C(x)$, e) magnetic field penetration with $\lambda^{-2}(T)$ linearly decreasing with $T$ for $T << T_C$,   f) $x$ dependent evolution of the isotope effect, g) a large condensation energy $\Delta\Omega(x)$, and an anomalously large discontinuity in the electron specific heat at $T_C(x_{op})$, h) an SC gap $\Delta_{\bf k}$ with d-wave symmetry, contrasting the s-wave gap in LTS materials, i) a tentative first order phase transition, with a concomitant quantum critical point.

In the normal state an anomalous property is the suppression of the density of electronic states referred to as the pseudogap.  The pseudogap $\Delta_{pg}(x)$, established by a number of spectroscopic probes, summarized in Ref. \cite{hufner}, deceases linearly in the underdoped domain from a maximum value $\Delta_{pg}(x \approx 0.05)$ until it merges on the overdoped side $x > x_{op}$ with the SC gap $\Delta(x)$.  Some researchers refer to $\Delta_{pg}$ as the SC gap, but there is considerable experimental evidence that the distinct lower energy SC gap exists, with a very different $x < x_{op}$ dependence.\cite{hufner,hashimoto}  Current theories either promote the pseudogap state as a pairing precursor of the SC state, or as a competitor due to unrelated dynamical fluctuations.\cite{trunin,kondo}  Hence, the origin and the effect of the pseudogap on the SC state remains unresolved.\cite{lee,hufner,norman,trunin}  The SPM gives new insight here.

Cuprate HTS theories abound.  Partially supported by experiment, various theories and reviews thereof include electron-phonon interactions\cite{gadermaier} with possible small polaron,\cite{alexandrov,zhou} or Jahn-Teller polaron formation,\cite{keller} interband interactions,\cite{bussmann, chen} exciton mediated interactions,\cite{geballe1, little, uemura, abrikosov, gorkov} negative U-center pairing,\cite{oganesyan, geballe2} bosonic electron-hole pairing,\cite{brinkman} spin-exciton,\cite{seradjeh} spin-phonon interactions,\cite{jarlborg} fractals,\cite{zaanen2} quantum oscillations in fermi liquids,\cite{chakravarty} quantum criticality,\cite{tallon1, marel, beyer, varma, balakirev} 2D strong electron-electron correlation with resonating valence bonds(RVB) and spin exchange, fluctuating spin exchange(overdoped regime), spinons, holons, various slave-particle techniques with extensive monte-carlo calculation, etc. \cite{anderson1,anderson3,fzhang,lee,monthoux,damascelli,kivelson,tsuei,chang,shimizu,dam,pathak}  The RVB theory,\cite{anderson2} generally implemented using the Hubbard $t-J$ Hamiltonian,\cite{anderson3} gives a reasonable perspective of the undoped antiferromagnetic charge transfer insulator phase, and as a potential HTS model, spin exchange has appeal since d-wave pairing is a natural consequence.\cite{anderson2}   Although various HTS characteristics are explained by reasonable, but disjointed, arguments,\cite{lee,emery} present model Hamiltonians produce only a small subset of the cuprate HTS properties a)-i).  For example, the $t-J$ Hamiltonian implementation of the RVB theory in the mean field(MF) approximation does not directly produce $T_C$, nor does it give the correct SC gap over the doping range of the SC state, nor does it give the observed ratio of the maximum pseudogap to the SC gap at optimal doping.\cite{anderson3}

The Bardeen-Cooper-Schrieffer(BCS) phonon mediated electron-pairing theory of superconductivity\cite{bcs,cooper}, in conjunction with the method of Eliashberg,\cite{eliashberg} applicable to the strong interaction regime, \cite{mcmillan} provides the framework for understanding the microscopic interactions responsible for LTS.  However, the BCS phonon mediated pairing interaction does not produce cuprate HTS, although the BCS theory appears to apply to MgB$_2$ with $T_C = 40^o$K.\cite{picket}   Multiple pairing interactions are considered necessary to explain cuprate HTS,\cite{lee, keller, little, geballe2, tsuei} but the exact nature of the doping dependent interactions in the cuprate unit cell presently remains beyond quantitative measurement.  Hence, formulation of a HTS model Hamiltonian relies on reasonable conjecture about the underlying mechanisms responsible for HTS, with subsequent validation requiring comparison of model predictions with many diverse experiments.

Our Hamiltonian $H(x)$ is based on a phonon-mediated interaction $V(x)$, detailed in Section III, and an exchange interaction $U(x)$, with the doping dependence $x$ determined by the SPM.  The $U(x) < 0$ is proportional an effective spin-singlet exchange, which is related to the $x$ dependent particle occupation probability of the O2p states.  The diagonal matrix elements of $U$ produce significant static fluctuation(deviation from the MF) even for relatively weak $U$.  We show that this static fluctuation is the key ingredient for HTS.  For weak $U$, the model predicts the listed a)-h) cuprate doping dependent characteristics with a second order phase transition(SOPT).  The SPM coupled with $H$ produces a unique relation between the conjoined model and the observed phase transition boundary $T_C(x)$, the SC gap $\Delta(T, x)$ and the NS pseudogap $\Delta_{pg}(x)$.  For stronger $U$, large static fluctuation results in an SC state that is essentially controlled by the ratio $U/V$.  If $U/V \lessapprox 1$ the model exhibits a first order phase transition(FOPT), and quantum criticality, a phenomenon of broad interest beyond cuprate HTS systems.\cite{coleman,kirchner}

\section{STATE PROBABILTIY MODEL}

Intrinsic cuprates are antiferromagnetic insulators with single or multiple CuO $xy$-planes, alternating with carrier reservoir planes which may have a significant effect on the value of $T_C(x_{op})$ in various cuprates.\cite{geballe2,suominen}   The Cu3d (Cu$^{2+}$) ion is in an octahedral environment surrounded by six O2p (O$^{2-}$) ions, with the apical oxygens on the $z$-axis.  Jahn-Teller distortion along the $z$-axis lowers the energy of the system by increasing the Cu apical O distance.  The resulting intrinsic cuprate is a quasi-2D antiferromagnetic insulator, exhibiting strong electron correlation.   In the $xy$-plane nearest neighbor Cu3d and O2p orbitals point directly toward each other, producing strong covalent bond coupling.  This intermediary ligand coupling indirectly connects neighboring Cu3d ions, giving rise to a large antiferromagnetic superexchange interaction $J_{dd} \sim 0.13$eV.

A salient property of cuprate HTS is its dependence on the doping concentration $x$.  For many cuprates, hole doping produces a dome shaped SC phase boundary in the doping range $[x_1 = 0.05, x_2 = 0.27]$ with $T_C(x_{op} \approx 0.16)$ exceeding $100^o$K; whereas for electron doping the range is comparatively narrow, $[x_1 = 0.14, x_2 < 0.2]$, with much lower $T_C(x_{op})$.  This section introduces a general state probability model based on particle occupation of the cuprate electronic states.  The SPM defines the probability of the SC state, and various normal states, in terms of the probabilities of the accessible unit cell(UC) states, independent of the model Hamiltonian.   The SPM mandates the existence plaquettes with net charge and spin to preserve local charge-spin neutrality.  Relating particle occupation probabilities to the doping concentration $x$, the SC gap $|\Delta(x)|$, the pseudogap $\Delta_{pg}(x)$, and a new anti-pseudogap $\Delta_{pg}'(x)$, which determine domain boundaries of the phase diagram, are a natural consequence of the SPM.  Linking the SPM to our Hamiltonian determines the doping dependence of the thermodynamic quantities.  Electron doping, with prediction of a possible new class of HTS materials, is considered at the end of this section.

Although there is considerable hybridization of the atomic states, it is advantageous to denote each UC state by the occupation of the constituent single particle states.   This simplification is consistent with a fundamental assumption of quantum mechanics that composite systems retain the properties of the individual constituent particles to a considerable extent.\cite{merzbacher}

The undoped cuprate state $\varphi_{AF}$ is charge neutral, Cu$^{2+}-$O$^{2-}$ with a hole on the Cu3d-orbitals and two electrons filling each of the O2 p$_x$ and p$_y$ orbitals.  For hole doping, the AF state, with few exceptions, \cite{lavrov} is rapidly quenched by doping induced hopping.\cite{lee}   In a doping range $0 \leq x < x_1$ there is an onset of charge transfer excitations involving random hopping of particles among Cu3d orbitals and O2p orbitals.  Initially holes are localized, but as doping increases hopping produces numerous possible states with net charge and spin.

The electronic state of a doped cuprate is modeled here by UC states containing a total of 5 fermion particles with spin $\sigma = \pm 1/2$.  These single-particle states form a large set of $2^5$ particle states $\times 2^5$ spin states $ = 2^{10} = 1024$ states denoted as $\varphi$.  It is advantageous to divide the UC states into two sets.  The sets $\varphi_{h\sigma}$ and $\varphi_{e\sigma}$ represent states with a hole or an electron, respectively, in a given Cu3d orbital.  The UC states, written as elements of a matrix, are
\begin{eqnarray}\label{states}
\varphi_{e\sigma}(ij) & \equiv & |e\sigma\rangle[\Phi\Phi^T]_{ij},\nonumber\\& &\\
\varphi_{h\sigma}(ij) & \equiv & |h\sigma\rangle[\Phi\Phi^T]_{ij},\nonumber\\& & \nonumber\\
\Phi^T & \equiv & [|e\sigma', h\sigma''\rangle,|h\sigma', e\sigma''\rangle, |h\sigma', h\sigma''\rangle, |e\sigma', e\sigma''\rangle].\nonumber
\end{eqnarray}
The single particle Cu3d states are denoted by $|e\sigma\rangle$ and $|h\sigma\rangle$.  Elements of the column vector $\Phi$, and the row vector $\Phi^T$ are states representing the possible particle configurations for the p$_y$-orbital, and the p$_x$-orbital, respectively.   The matrix $\Phi\Phi^T$ gives the complete set of particle configurations, for each spin set.  The $\varphi_{h\sigma}$ states are illustrated in table \ref{hstates}.
 \begin{table}[t]
 \renewcommand{\arraystretch}{1.1}
 \protect\setlength{\arraycolsep}{3pt}
\[\begin{array}{llll llll llll lll}
\bullet&&&&\circ&&&&\bullet&&&&\circ\\			
\bullet&&&&\bullet&&&&\circ&&&&\circ\\	
\circ&\bullet&\bullet&\quad\quad &\circ&\bullet&\bullet&\quad\quad &\circ&\bullet&\bullet&\quad\quad &\circ&\bullet&\bullet\\
\\
\bullet&&&&\circ&&&&\bullet&&&&\circ\\
\bullet&&&&\bullet&&&&\circ&&&&\circ\\			
\circ&\bullet&\circ&\quad\quad &\circ&\bullet&\circ&\quad\quad &\circ&\bullet&\circ&\quad\quad &\circ&\bullet&\circ\\
\\
\bullet&&&&\circ&&&&\bullet&&&&\circ\\
\bullet&&&&\bullet&&&&\circ&&&&\circ\\			
\circ&\circ&\bullet&\quad\quad &\circ&\circ&\bullet&\quad\quad &\circ&\circ&\bullet&\quad\quad &\circ&\circ&\bullet\\
\\
\bullet&&&&\circ&&&&\bullet&&&&\circ\\
\bullet&&&&\bullet&&&&\circ&&&&\circ\\			
\circ&\circ&\circ&\quad\quad &\circ&\circ&\circ&\quad\quad &\circ&\circ&\circ&\quad\quad &\circ&\circ&\circ
\end{array}\]
\caption{\label{hstates} Illustrated are the distinct single-particle states comprising the unit cell states $\varphi_{h\sigma}(ij)$.  Holes are denoted by $\circ$, and electrons by $\bullet$, with spin not shown.  The corner position represents one of the Cu3d orbitals.  The undoped, charge neutral AF state is $\varphi(11)$.  Other states have net charge, e.g. $\varphi(12)$ has charge $+e$, $\varphi(22)$ has $+2e$, and $\varphi(24)$ has $+3e$.  The UC states $\varphi_{e\sigma}(ij)$ are obtained by interchanging every $\bullet$ and $\circ$, with the same spin.  The net charge of $\varphi_{e\sigma}(ij)$ and $\varphi_{h\sigma}(ij)$ differ.}
\end{table}

Contrasting the well defined undoped AF state, the doped state is a probabilistic mixture of the states in Eq. (\ref{states}) that result from doping induced hopping.  The constituent single particle states are assumed to be mutually exclusive, and thus the distinct UC states are mutually exclusive, analogous to a 16 sided die with each face imprinted with one of the particle configurations, including spin.  Let $P_{e\sigma}(x, T)$ represent the energy averaged probability that a given single particle state is occupied by an electron with spin $\sigma$, at temperature $T$, and doping concentration $x$.  The corresponding hole occupation probability is $P_{h\sigma} = 1 - P_{e\sigma}$.  The probability that a particular UC state $\varphi_{h\sigma}(ij)$ exists is a joint probability involving products of the probabilities $P_{e\sigma}$, and $P_{h\sigma}$.  With one high $T$ exception in the normal state, the scaled properties developed below have negligible $T$ dependence.

The states $\varphi_{h\sigma}(ij)$ and $\varphi_{e\sigma}(ij)$ carry net charge-spin.  To maintain charge-spin neutrality in the CuO plane, it is energetically favorable for plaquettes containing UC states to form with opposite charge and spin for each value of $x$.  (Plaquette is used in a generic sense to include various geometrical shapes commensurate with lattice symmetry.)  The sets of plaquette states necessary to preserve charge-spin neutrality, denoted as
\begin{equation}\label{plaqstates}
\varphi_{h\uparrow},\quad \varphi_{e\uparrow},\quad \varphi_{h\downarrow},\quad \varphi_{e\downarrow},
\end{equation}
must satisfy the state probability equalities
\begin{equation}\label{Probplaqstates}
P(\varphi_{h\uparrow}) = P(\varphi_{e\downarrow}),\quad P(\varphi_{e\uparrow}) = P(\varphi_{h\downarrow}),
\end{equation}
where $P(\varphi)$ is the probability that $\varphi$ exists.  Having established the UC states and the probability requirement Eq. (\ref{Probplaqstates}) for pairing the plaquettes, one can determine SC and normal state probabilities and average properties.

\textbf{SC State:} In a doping range $x_1 < x < x_2$ a coherent doping dependent SC state $\varphi_{SC}$ emerges for $T \leq T_C$.  An effective exchange interaction responsible for the SC state is formulated below.  Based on repulsive energy considerations, it is assumed that the SC state excludes the AF state $\varphi(11)$ and the extreme overdoped state $\varphi(44)$, with completely filled or completely empty p-orbitals, respectively.  In the overdoped range $x \approx x_2$ states form in the charge transfer gap, resulting in a Fermi-liquid state $\varphi_{FL}$ with negligible remaining individual particle character, and the SC state is destroyed.

We define the probability that the SC state exists as the sum of the probabilities of the UC states $\varphi_h\sigma$ that remain after extracting the high energy states $\varphi(11)$ and $\varphi(44)$ from Table \ref{hstates}, and the corresponding $\varphi_e\sigma$ states.  The result is
\begin{eqnarray}\label{ProbSCeh}
P_{SC}(\varphi_{e\sigma}) & = & P_{e\sigma}F_p,\quad P_{SC}(\varphi_{h\sigma}) = P_{h\sigma}F_p \\&&\nonumber\\
F_p & = & (P_{e\sigma'} + P_{e\sigma''})(P_{h\sigma'} + P_{h\sigma''}] - \nonumber\\&&\nonumber\\
   & & 2P_{e\sigma'}P_{e\sigma''}P_{h\sigma'}P_{h\sigma''}.\nonumber
\end{eqnarray}
Adding the expressions in Eq. (\ref{ProbSCeh}) gives the SC state probability $P_{SC} = F_p$, which depends only on the occupation probabilities for the p-orbitals.  For the remainder of the analysis of the SC state we neglect spin dependence of the p-orbital probabilities $P_{e\sigma'}$ and  $P_{h\sigma'}$, giving
\begin{equation}\label{ProbSC1}
P_{SC} = 4P_eP_h[1 - \frac{1}{2}P_eP_h].
\end{equation}
The first term in $P_{SC}$ arises from a simplified three particle state model by considering the row and columns in table \ref{hstates} independently, excluding doubly occupied $p$-orbitals.  The second term results from a joint probabilty involving both p$_x$ and p$_y$-orbitals.   Thus $P_{SC}$, as defined above, automatically excludes all doubly occupied p-orbitals.  Writing $P_{SC}$ in the form $P_{SC}  =  1 - P_e^4 - P_h^4$ confirms that $\varphi(11)$ and $\varphi(44)$ are excluded from the SC state.

The utility of $P_{SC}$ is implemented by relating the particle occupation probabilities to the doping concentration $x$.  Assuming a uniform probability density over a doping range $w = x_2 - x_1$, the probability that a doped hole is created in doping range $[x_1, x_1 + x]$ is $\Pi_h(x) = (x-x_1)/w$.  (Setting $d\Pi/dx$ constant, and neglecting temperature dependence, is the same assumption used in Anderson Ref. \cite{anderson3}, but its application and the resultant HTS properties are clearly not the same, as discussed below.)  The doped holes fill oxygen orbitals in the reservoir planes, which in turn distort the lattice and induce particle hopping in the CuO conducting planes.   Since the probability of the SC state, Eq. (\ref{ProbSC1}), reduced to the that of effective p orbital states with an electron and a hole, it is reasonable to assume that the probability $P_h$ that a position labeled $h$ has a hole is $P_h \propto \Pi_h(x)$.  Without loss of essential information, we set the constant of proportionality to unity, giving the probability $P _e = \Pi_e(x) = 1 - \Pi_h(x)$ that a position labeled $e$ has an electron.   Hence, Eq. (\ref{ProbSC1}) written in terms of the dopant probabilities is
\begin{eqnarray}\label{ProbSC2}
P_{SC}(x) & = & 4\Pi_e(x)\Pi_h(x)[1 - \frac{1}{2}\Pi_e(x)\Pi_h(x)],\\&&\nonumber\\
\Pi_h(x) & = & \frac{x - x_1}{w},\quad \Pi_e(x) = \frac{x_2 - x}{w},\quad w = x_2 - x_1,\nonumber
\end{eqnarray}
with $\Pi_e(x) + \Pi_h(x) = 1$.

Equations (\ref{ProbSCeh})-(\ref{ProbSC2}) are independent of the detailed interaction responsible for the SC state.   The proposed pairing interaction is a net effective inter-plaquette, spin singlet exchange $J(x)$ per unit cell that emerges as a consequence of doping induced hopping.  The $J(x)$ is an average over the values for each UC state $\varphi_{ij}$ participating in the SC state.  Determination of the exchange $J_{ij}$ between states $\varphi_{h\uparrow}(ij)$ and $\varphi_{e\downarrow}(ij)$ requires microscopic analysis to determine the overlap integrals and the resultant eigenstates of a very complicated system.  Here, the relative value $J(x)/J(x_{op})$ is obtained by probability arguments, with $J(x_{op})$ a parameter found by fitting experimental data in Section V.

The exchange energy, per unit cell, averaged over the accessible UC state configurations $\varphi_{i,j}$ is formulated in Appendix A.  The result from Eq. (\ref{J}) is
\begin{eqnarray}\label{Jx}
J(x)& \approx & J_2\Pi_e(x)\Pi_h(x)[1 - 2(1 - \frac{J_1}{2J_2})\Pi_e\Pi_h] \\&&\nonumber\\
J_1 & = & J_{11} + J_{22} + 2(J_{12} + J_{34}),\nonumber\\&&\nonumber\\
J_2 & = & 2(J_{13} + J_{23}), \quad J_3 = 2(J_{14} + J_{24}),\nonumber
\end{eqnarray}
The $J_{ij}$ are the exchange constants corresponding to the interaction between $\varphi_{h\uparrow}(i,j)$ and $\varphi_{e\downarrow}(i,j)$.  Thus $J(x)$ is an average spin-singlet exchange between the paired plaquettes that ensure local charge-spin neutrality.  Although microscopic evaluation of the $J_{ij}$ involves multiple particle-particle and particle-antiparticle interactions between UC states, the doping dependence of $J(x)$ is via the $\Pi(x)$'s which refer to p orbital occupation.  The participation of the Cu3d particle serves as a mediator, appearing only in the exchange constants.  Note that for $J_1/J_ 2= 3/2$, the effective exchange is identical to $P_{SC}$, to within a constant of proportionality.   Since the maximum of $(1/2)\Pi_e(x)\Pi_h(x) = 1/8$ the small $p_x$-$p_y$ interaction correction will be neglected below.  In this approximation the $J_{ij}$ in the factor $J_2$ have no effect on scaled HTS properties.

Retaining the linear two particle terms in Eqs. (\ref{ProbSC2}) and (\ref{Jx}), the effective exchange $J(x)$ and the SC state probabilities are given by
\begin{equation}\label{JProbSC}
J(x)  \propto  P_{SC}(x) =  4\Pi_e(x)\Pi_h(x).
\end{equation}
The simplification to the dominant two particle term implies that an effective electron-hole pair within the p$_x$ and p$_y$ orbitals is an essential ingredient characterizing the SC state.  Assuming a lower energy alternating spin state, the implied quasi-particles are spin-excitons with spin-singlet exchange.  This picture is somewhat distinct from the Zhang-Rice\cite{fzhang} and Geballe\cite{geballe2} pictures which explicitly contain a Cu3d particle.  All three scenarios involve dynamical processes, which are treated here as static, time-averaged, phenomena.

In view of Eq. (\ref{JProbSC}), we propose an hypothesis:
\begin{quote}
\emph{Doping dependence of scaled energy parameters that characterize SC, and normal, states is manifested only via the doping dependent probabilities that the relevant UC states are accessible.}
\end{quote}
The validity of this hypothesis, already evident in Eq. (\ref{JProbSC}), is further substantiated by the following analysis that produces the doping dependence of many observed cuprate properties.

In accordance with the hypothesis, the scaled SC gap $\Delta(x) \propto \Pi_h(x)\Pi_e(x)$.   This result is also confirmed by combining Eq. (\ref{linear}), which is a direct consequence our model Hamiltonian developed below, with Eq. (\ref{JProbSC}).  This gives the doping dependent relations
\begin{equation}\label{universe}
\frac{T_C(x)}{T_C(x_{op})} = \frac{\Delta(x)}{\Delta_0} = 4\Pi_h(x)\Pi_e(x), \quad \Delta_0 = \Delta(x_{op}).
\end{equation}
These universal relations, independent of the average density of states $N_0$ and the cutoff temperature $T_m$ defined in Section IV, give the doping dependent phase boundary $T_C(x)$ and the low temperature SC gap $\Delta(x)$ observed in cuprates, as shown in Section V.  Equation (\ref{universe}), concomitant with the Hamiltonian model relation $T_C(\nu)$, is supported by the $(T, x)$ dependence of the Hall-coefficient used to track the hole(electron) charge characteristic throughout the SC phase.\cite{brinkman}  The two-particle nature of the pairing interaction parameter $\nu(x)$ and the resulting $\Delta(x)$, is consistent with measurement of the SC gap, requiring two-particle probes.\cite{hufner}  In the RVB-Hubbard model, Ref. \cite{anderson3}, the SC gap is assumed to be proportional to $g_t^2(x) = [2x/(1+x)]^2$, where $g_t$ is a kinetic energy renormalization factor.  This function only approximates $\Delta(x)$ in the very underdoped region $x << x_{op}$, whereas Eq. (\ref{universe}) agrees with the observed gap over the entire doping range of the SC state.

\textbf{Normal State:} The doping dependent normal state is characterized by unusual properties, e.g. pseudogap, vortices, stripes, etc, reviewed in Ref. \cite{lee}.   The apparent complexity is daunting, as are the myriad of complex and exotic theories.  Nichtsdestoweniger, it is our contention that there is a rather simple explanation for many properties based on the UC states in Eq. (\ref{states}) and illustrated in Table \ref{hstates}.

Since doping induced hopping is a random process, it is asserted that the normal state is characterized by the complete set of UC states in Eq. (\ref{plaqstates}).  Assuming that all of the state configurations in Eq. (\ref{states}) are accessible, the probability $P(\varphi_h)$ that some one of the states in $\varphi_h$ exists is denoted by
\begin{equation}\label{ProbPhie}
P(\varphi_{h\sigma}) = P(|h\sigma\rangle\Phi\Phi^T) = P_{h\sigma}P(\Phi)P(\Phi^T).
\end{equation}
The probability that $\Phi$ exists is defined as the sum of the probabilities for each state, i.e. it is the probability that some one of the states exist.  Thus
\begin{equation}\label{ProbPhi}
P(\Phi)  =  (P_{e\sigma'} + P_{h\sigma'})(P_{e\sigma''} + P_{h\sigma''}) = 1.
\end{equation}
Since $P(\Phi)^T$ is also unity, one obtains
\begin{equation}\label{Probvarphi}
P(\varphi_{h\sigma}) = P_{h\sigma}(x, T),\quad P(\varphi_{e\sigma}) = P_{e\sigma}(x, T),
\end{equation}
independent of the p-orbital occupation for any given spin set.  One can derive the same result for $P(\varphi_{h\sigma})$ by tediously summing the probabilities for each of the UC states $\varphi_{h\sigma}$ in Table \ref{hstates}, and for the corresponding $\varphi_{e\sigma}$ states.

The seemingly trivial expression in Eq. (\ref{Probvarphi}), a consequence of the completeness of the UC p-orbital states used, has profound implications.  It gives a non-zero probability for static charge and spin fluctuation on any given Cu3d orbital.  Since $P_h(x) \neq P_e(x)$ for $x \neq x_{op}$, it is evident that the states $\varphi_h$ and $\varphi_e$ cannot exist alone.  As concluded above, plaquettes form in the CuO plane with opposite charge and spin for each value of $x$.  The presence of such plaquettes is consistent with the formation of charge and/or spin density waves with concomitant gaps.\cite{lee,callaway}  The paired plaquettes dictate that the system is composed of the four states in Eq. (\ref{plaqstates}).  Since Eq. (\ref{Probvarphi}) shows that the doping dependence of these states is characterized by the Cu3d orbital occupation, we set $P(x) = \Pi(x)$, giving the plaquette state probabilities

\begin{eqnarray}\label{Probpm}
P(\varphi_{h\uparrow}) & = & \Pi_h(x),\quad P(\varphi_{e\uparrow}) = \Pi_e(x), \nonumber\\&&\\
P(\varphi_{h\downarrow}) & = & \Pi_e(x),\quad P(\varphi_{e\downarrow}) = \Pi_h(x).\nonumber
\end{eqnarray}
Consider a checkerboard pattern of UC plaquette states $\varphi_{\uparrow}$, and $\varphi_{\downarrow}$.  Since $\Pi_e(x) + \Pi_h(x) = 1$, the plaquette state $\varphi_{\uparrow}(x < x_{op})$ is dominated by spin-up electrons, and the plaquette state $\varphi_{\downarrow}(x < x_{op})$ is dominated by spin-down holes.  At optimal doping $x = x_{op}$ there is no net charge or spin for the plaquette pair.  Overdoping gives the reverse of the underdoped picture.

Since the probability of the normal states reduced to simple, one particle probabilities for the Cu3d orbitals, we are able to formulate the exchange interaction between plaquettes in the normal states more precisely than that in the SC state.  Let $ J_0 = J(h\uparrow,e\downarrow) = - J(e\uparrow,h\downarrow)$ denote the spin singlet exchange between spins in the plaquette states characterized by the corresponding particle and spin in the Cu3d orbital. Neglecting particle-particle exchange, expected to be much weaker than particle-antiparticle exchange, and noting that exchange requires joint probabilities, the doping dependent average interaction energy $J_s(x)$ for the balanced plaquette pair is
\begin{equation}\label{Ex}
J_s(x) = J_0[\Pi_h^2(x) - \Pi_e^2(x)] = J_0[\Pi_h(x) - \Pi_e(x)].
\end{equation}
Introducing the definitions
\begin{eqnarray}\label{Deltacharge}
\Delta_h(x) & = & J_0 + J_s(x) = 2J_0\Pi_h(x)\nonumber\\&&\\
\Delta_e(x) & = & J_0 - J_s(x) = 2J_0\Pi_e(x),\nonumber
\end{eqnarray}
the energy $\Delta_e(x)$ becomes identical to the experimental pseudogap $\Delta_{pg}(x)$ with the single requirement that $\Delta_e(x) \geq \Delta(x)$.  This gives the maximum SC gap $\Delta_0 = 0.5J_0$ such that $\Delta_e(x)$ is excluded from the SC state, yielding the form
\begin{equation}\label{pg}
\Delta_{pg}(x) \equiv \Delta_e(x) = 4\Delta_0\Pi_e(x).
\end{equation}
The pseudogap $\Delta_{pg}(x)$ in Eq. (\ref{pg}) decreases linearly with doping from a maximum $\Delta_{pg}(x_1) = 4\Delta_0$ to $\Delta_{pg}(x_2) = \Delta(x_2) = 0$.   It is shown in Section V that both Eqs. (\ref{universe}) and (\ref{pg}) are in excellent agreement with a broad class of cuprates.  In comparison, the RVB-Hubbard theory in Ref. \cite{anderson3}, Figure 2, also gives $\Delta_{pg}(x) \geq \Delta(x)$, but $\Delta_{pg}(x_1) > 6\Delta_0$ does not agree with experiment.

A simple SC gap-pseudogap relation is obtained by eliminating the scaling factor $\Delta_0$ from Eqs. (\ref{universe}) and (\ref{pg}), or alternatively retaining $\Delta_0$ and writing a difference relation.  This gives
\begin{equation}\label{scpg}
\Delta(x) = \Delta_{pg}(x)\Pi_h(x) = \Delta_{pg}(x) - 4\Delta_0\Pi_{e}^2(x),
\end{equation}
which is a universal relation corresponding to Eq. (\ref{universe}).  It is evident that $\Delta_{pg}$ and $\Delta$ have the same symmetry, and that their difference is proportional to the probability for filled $p$ orbitals, which characterize the AF state.

It is useful to introduce the concept of a pseudogap state, and an anti-pseudogap state, defined by their probabilities $P_{pg}(x) = \Pi_e(x)$, and $P_{pg}^{\prime}(x) = \Pi_h(x)$, respectively, corresponding to the pseudogap $\Delta_{pg}(x) = \Delta_e(x)$ and anti-pseudogap $\Delta_{pg}^{\prime}(x) = \Delta_h(x)$.   The transition domains between the pseudogap(anti-pseudogap) state and the SC state is defined by the joint probabilities
\begin{equation}\label{Probtrans}
P_{tr}(x) = P_{pg}(x)P_{SC}(x),\quad P_{tr}^{\prime}(x) = P_{pg}^{\prime}(x)P_{SC}(x)
\end{equation}
where  $P_{SC}(x)$ is given by Eq. (\ref{ProbSC2}).  The corresponding energy transition boundaries are $\Delta_{tr}(x) = \Delta_0P_{tr}(x)$ and $\Delta_{tr}^{\prime}(x) = \Delta_0P_{tr}^{\prime}(x)$.

\begin{figure}
\includegraphics[width = 3.2in]{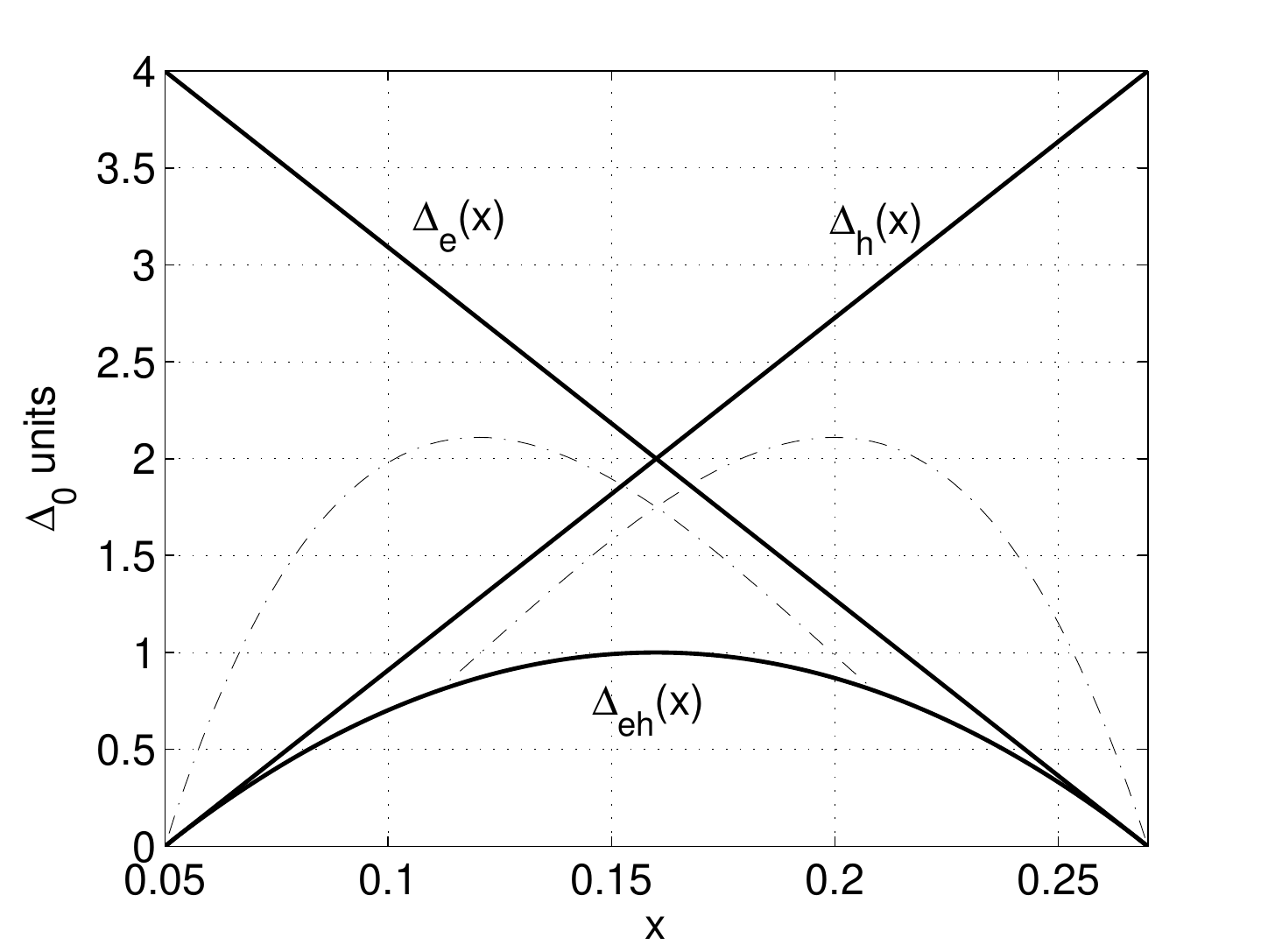}
\caption{\label{Fig1} Plotted is the doping concentration $x$ dependence of the common, two-particle SC gap $\Delta_{eh}(x) \equiv \Delta (x)$ and the distinct, single-particle pseudogaps $\Delta_e(x)$ and $\Delta_h(x)$ for hole doped cuprates and the proposed electron doped anti-cuprates, respectively.  The dashed curves are the transition energies $\Delta_{tr}(x)$ and $\Delta_{tr}^{\prime}(x)$.}
\end{figure}

The SPM exhibits reflection symmetry about $x_{op}$ for energies in both the SC and normal states.  The symmetry is illustrated in Fig.\ref{Fig1}, where the SC two-particle gap $\Delta(x) = \Delta_{eh}(x)$ and the distinct, single-particle, pseudo gaps $\Delta_e(x)$ and $\Delta_h(x)$ are plotted in units of $\Delta_0$ versus the doping concentration $x$.  The dashed curves are $\Delta_{tr}(x)$ and $\Delta_{tr}^{\prime}(x)$, representing the boundary of the transition from the normal state to the SC state.  Not shown, but of interest, is the net normal state pseudogap energy $E_{pg}(x)= (1/4\Delta_0)[\Delta_e^2(x) + \Delta_h^2(x)$].  As $x$ increases on the underdoped side $E_{pg}(x)$ and the pseudogap decrease and the SC state becomes more robust.  The SC state is maximized at optimal doping where $E_{pg}(x)$ is minimum, with equal pseudogaps and no net plaquette-plaquette interaction energy in the normal state.  On the overdoped side $E_{pg}(x)$ and the anti-pseudogap increases and the SC state loses coherence.

A pertinent question is why is the pseudogap $\Delta_e(x)$ observed, while $\Delta_h(x)$ has not yet been observed.   In the overdoped range, it is plausible that a local excess hole imbalance would be obscured, or destroyed, by the onset of the fermi-fluid hole state, but $\Delta_h(x)$ should appear close to the SC gap on the underdoped side near $x_{op}$.  In any case, $\Delta_h(x)$ has a significant role in both cuprates and the anti-cuprates proposed below.

The doping dependence of the Knight-shift, which is proportional to the spin susceptibility $\chi_s$, is partially explained by $\Delta_e(x, T)$.  Here the width of the probability distribution $w(T) = x_2(T) - x_1(T)$ is assumed to be a function of $T$, with $x_{op} = (x_1 + x_2)/2$ constant.  Since $\Delta_e(x, T)$ is proportional to the spin-singlet exchange between plaquettes, $\chi_s(x, T) \propto 1/\Delta_e(x, T)$, giving

\begin{equation}\label{chis}
\frac{\chi_s(x_{op}, T)}{\chi_s(x, T)} = \frac{\Delta_e(x_{op})}{\Delta_e(x)} = 1 + \frac{2}{w(T)}(x_{op} - x)
\end{equation}
In the underdoped range $\chi_s(x, T) < \chi_s(x_{op}, T)$, and in the overdoped range $\chi_s(x, T) > \chi_s(x_{op}, T)$ for each value of $T$ as observed in La$_{2-x}$Sr$_x$CuO$_2$.\cite{nakano}  The $T$ dependence of $w(T)$ is not obtained within the SPM.  However, using a Curie-Weiss form $w(T) = T - \Theta$ approximates the $\chi_s(x, T)$ curves.

For hole doping of cuprates, a few more comparisons are noteworthy.  The slave-boson picture developed from the $t-J$ model,\cite{lee} divides the phase diagram into domains by a fermi pairing parameter and a MF bose condensation parameter, which have the same doping dependence as $\Delta_e(x)$ and $\Delta_h(x)$, respectively.  The Nernst domain is very similar to our pseudogap-SC state transition domain, bounded by $\Delta_{tr}(x)$ and the SC gap $\Delta(x)$.

In the Emery-Kivelson model\cite{emery} $x_{op}$ is determined by the intersection of a pairing amplitude $\Delta_0(x)$, which decreases with $x$, and the phase stiffness $E_{\theta}(x) = k_BT_{\theta}$, which increases with $x$.  The value of $T_C(x_{op})$ is generally not the maximum value of $T_C(x)$.\cite{geballe2}  In our model, the intersection of the pseudogap $\Delta_e(x)$ and the anti-pseudogap $\Delta_h(x)$ occurs at $x_{op}$, with $T_C(x_{op})$ the maximum value.

Measurements of the superfluid density $n_s(T, x)$\cite{trunin} provide another connection to our model.  The $x$ dependence $n_s(T = 0, x) \propto x - x_1$ is the same as the that of anti-pseudo gap $\Delta_{pg}'(x) \propto \Pi_h(x)$.   Also, $dn_s(T, x)/dT \propto \Pi_e^2(x)$, which is the probability that a $p$ orbital is full, fits the data as well as the singular function $dn_s(T, x)/dT \propto x^{-2}$ proposed.\cite{trunin}

\textbf{Electron Doping and Anti-Cuprates:}  Electron doping of cuprates produces an SC state in the comparatively narrow range, $[x_1 = 0.14, x_2 < 0.2]$, with much lower $T_C(x_{op})$ than for hole doping.  Referring to the discussion above, the SC state of cuprates stems from the states in Table \ref{hstates}, excluding $\varphi(11)$ and $\varphi(44)$.  Hole doping increases the probability that these states exist at the expense of the definite, undoped AF state $\varphi_h(11)$.  In contrast, while electron doping may induce hopping, it also serves to maintain the predominance of the undoped state to higher values of $x$, thus suppressing the formation of the states necessary to form the SC state.  At a doping induced hopping level which finally produces an SC state, the onset of an electron Fermi-fluid state $\varphi_{FL}$ again quickly suppresses the SC state.

Electron doping of cuprates is not a very effective means of producing HTS.  However, envision a possible class of "anti-cuprates" with the same crystal structure as a cuprate, but with exchanged anion and cation roles: The Cu$^{2+}$ and the O$^{2-}$ are replaced by ions A$^{2-}$ and B$^{2+}$, respectively.   The SC gap is the same as that for cuprates, but the role of the pseudogap is now taken by the anti-pseudogap $\Delta_h(x)$.   If anti-cuprate material exists, or can be synthesized, electron doping should exhibit HTS properties similar to hole doping of cuprates.  The symmetry of the two HTS systems is evident in Fig.\ref{Fig1}.

\section{Hamiltonian $H$}

The Hamiltonian is constructed from a combined interaction $V + U$.  As analyzed in Section II, a consequence of particle hopping is an induced effective spin-singlet exchange interaction $U(x) \propto J(x)$.  However, it is emphasized that the development in this section, and many of the resulting HTS properties, are completely independent of the $x$ dependence of $U$.  In addition to $U$, the formation of small polarons contribute to a phonon-mediated interaction $V$.  The resulting two-particle interaction $H_{int}$ has the general form
\begin{eqnarray}\label{Hintq}
H_{int} & = & \frac{1}{N}\sum_{\bf q}\sum_{\bf k \bf {k}^\prime}\Gamma_{\bf k \bf {k}^\prime}p_{\bf k q}^{\dagger} p_{{\bf k}^{\prime}\bf q},\\& &\nonumber \\
p_{\bf k q} & = & c_{-{\bf k} + {\bf q}/2\downarrow}c_{{\bf k} + {\bf q}/2\uparrow},\nonumber
\end{eqnarray}
where the interaction matrix ${\bf \Gamma} = [\Gamma_{{\bf kk}^\prime}] = [V_{{\bf kk}^{\prime}} + U_{{\bf kk}^{\prime}}]$.

In the cuprate SC state it is consistent with observation that BCS pairs $c_{-\bf k \downarrow}c_{\bf k \uparrow}$ form in the CuO planes.\cite{gough,tinkham}    Accordingly, $H_{int}$ is approximated by replacing all terms in the operator products in Eq. (\ref{Hintq}) with BCS pairing terms, i.e. neglecting ${\bf q}$ dependence.  The Hamiltonian for the electron system, including the kinetic energy is
\begin{eqnarray}\label{H1}
H & = & H_{kin} + H_{int},\\& &\nonumber \\
H_{kin} & = & \sum_{\bf k } \varepsilon_{\bf k}[c_{\bf k}^{\dagger}c_{\bf k } + c_{-\bf k }^{\dagger}c_{-\bf k}],\nonumber\\ & & \nonumber\\
H_{int}({\bf \Gamma}) & = & {\bf p}^{\dagger}{\bf \Gamma} {\bf p} = \sum_{\bf k \bf {k}^\prime} \Gamma_{\bf k \bf {k}^\prime } (c_{-\bf k}c_{\bf k})^{\dagger} c_{-\bf {k}^\prime }c_{\bf {k}^\prime },\nonumber
\end{eqnarray}
where ${\bf p} = [c_{-\bf k}c_{\bf k}]$ is a column vector with BCS pairing operators as the elements.  The fixed spins are implicitly indicated by $\pm\bf k \equiv {\bf k}\uparrow , -{\bf k}\downarrow$.   The single particle energies, referenced to the chemical potential $\mu$, are $\varepsilon_{k} = \epsilon_{\bf k} - \mu $.  In the underdoped regime $\epsilon_{\bf k}$ is the tight binding kinetic energy, and in the overdoped regime it is expected that $\epsilon_{\bf k}$ is characterized by an effective mass $m^*$.  The general form of $\bf \Gamma$ is retained in the analysis below until it becomes necessary to develop the specific matrix structure relevant to cuprates.  Next, a method is introduced for dealing with non-negligible static fluctuation due to $\Gamma_{\bf kk} \neq 0$.

\textbf{Mean Field and Static Fluctuation:} Our procedure begins with an extraction of the mean field(MF) part $H_{mf}$ of $H$ leaving a deviation(static fluctuation) $H_d$.  Then $H_{mf}$ is diagonalized and $H_d$ is approximated by $\langle H_d\rangle$, which is subsequently evaluated exactly in the eigenstates of $H_{mf}$.  The first step is to introduce a vector pair deviation operator ${\bf d} = [d_{\bf k}] = {\bf b} - {\bf p}$, where ${\bf d} = [d_{\bf k}]$ and the components of ${\bf b} = [b_{\bf k}]$ are complex scalar fields $b_{\bf k }(x)$.  Using $\bf d$ to reorganize $H_{int}({\bf \Gamma})$ in Eq. (\ref{H1}) gives the form
\begin{eqnarray}\label{H2}
H & = & H_{kin} + H_b({\bf \Gamma}) + H_{d}({\bf \Gamma}),\\&  &\nonumber\\
H_b & = & \sum_{\bf k \bf {k}^\prime }\Gamma_{\bf k \bf {k}^\prime} [b_{\bf k}^* c_{-\bf {k}^\prime }c_{\bf {k}^\prime } + (c_{-\bf k}c_{\bf k})^{\dagger}b_{\bf {k}^\prime} - b_{\bf k}^*b_{\bf {k}^\prime}],\nonumber\\&{}\nonumber &\\
H_{d} & = & \sum_{\bf k \bf {k}^\prime}\Gamma_{\bf k \bf {k}^\prime}d_{\bf k}^{\dagger}d_{\bf {k}^\prime}.\nonumber
\end{eqnarray}
If $H_d$ is neglected, $H$ reduces to the form of the conventional BCS MF Hamiltonian with $b_{\bf k } = \langle c_{-\bf k \downarrow}c_{\bf k \uparrow}\rangle$.\cite{tinkham, callaway}  The MF approximation is valid for interactions with only off-diagonal matrix elements, such as the often used model with $\Gamma_{\bf k \bf {k}^\prime} = - V_0(1 - \delta_{{\bf kk}'})$.  In this case, static fluctuations are zero in the random phase approximation and $H_d$ is negligible.  However, as shown below, $H_d$ is not negligible for a pairing interaction with diagonal elements $\Gamma_{\bf kk} \neq 0$.  This key feature of $\bf \Gamma$ produces HTS.

The Hamiltonian in Eq. (\ref{H2}) is a reformulation of $H$ in Eq. (\ref{H1}), with no approximation.  The Bogoliubov-Valatin canonical transformation\cite{bogoliubov,valatin} diagonalizes $H _{mf} = H_{kin} + H_b({\bf \Gamma})$ in terms of quasi-particle number operators $n_{\bf k} = \gamma_{\bf k}^{\dagger}\gamma_{\bf k}$, and $q_{\bf k} = \lambda_{\bf k}^{\dagger} \lambda_{\bf k}$.  To complete the diagonalization of $H$, the deviation term $H_d$ is approximated by its ensemble average $\langle H_d\rangle = \langle{\bf d}^{\dagger}{\bf \Gamma}{\bf d}\rangle$.  The details are outlined in Appendix A.  Exact evaluation of $\langle d_{\bf k}^{\dagger}d_{\bf {k}^\prime}\rangle $ in the eigenstates of $H_{mf}$, using Eqs. (\ref{Hmf}) and (\ref{dd}), gives the diagonal Hamiltonian
\begin{eqnarray}\label{H}
H & = &  H_{mf} + \langle H_d\rangle,\\& &\nonumber\\
H_{mf} &= & -\sum_{\bf k}[E_{\bf k }(1-n_{\bf k} - q_{\bf k}) - \varepsilon_{\bf k} - \Delta_{\bf k}b_{\bf k}^*],\nonumber\\&{}\nonumber &\\
\langle H_d\rangle & = & \sum_{\bf k \bf {k}^\prime}\Gamma_{\bf k \bf {k}^\prime}[\langle d_{\bf k}^{\dagger}\rangle\langle d_{\bf k^\prime}\rangle + \frac{1}{4}\delta_{\bf k \bf {k}^{\prime}}\sigma_{\bf k}^2], \nonumber \\&{}\nonumber &\\
\sigma_{\bf k} & = & 1 - \frac{\varepsilon_{\bf k}}{E_{\bf k }}\tanh(\beta E_{\bf k }/2),~~~\beta = 1/(k_BT), \nonumber\\&{}\nonumber &\\
E_{\bf k}& = & \sqrt{\varepsilon_{\bf k}^2 + |\Delta_{\bf k}|^2}.\nonumber
\end{eqnarray}
The quasi-particle excitation energy $E_{\bf k}$ depends on the energy gap $\Delta_{\bf k}$, which is linked to the average gap deviation $\langle \delta_{\bf k}\rangle$.  They are defined by
\begin{equation}\label{Deltak}
\Delta_{\bf k} = - \sum_{\bf k^\prime}\Gamma_{\bf k \bf {k}^\prime} b_{\bf k^\prime},\quad \langle\delta_{\bf k}\rangle = -\sum_{\bf {k}^\prime}\Gamma_{\bf k \bf {k}^\prime}\langle d_{\bf k ^\prime}\rangle .
\end{equation}
Using $\langle d_{\bf k }\rangle = b_{\bf k} - \langle c_{-\bf k }c_{\bf k}\rangle$, and Eq. (\ref{pairtf}) to evaluate $\langle c_{-\bf k }c_{\bf k}\rangle$, the gap $\Delta_{\bf k}$ and gap deviation $\langle \delta_{\bf k}\rangle$ in Eq. (\ref{Deltak}) are related by the constraint
\begin{eqnarray}\label{Ddconk}
\frac{\langle \delta_{\bf k}\rangle}{\Delta_{\bf k}} & = & g_{\bf k} = 1  + \frac{1}{\Delta_{\bf k}}\sum_{\bf k ^\prime} \Gamma_{\bf k \bf {k}^\prime}\langle c_{-\bf {k}^\prime }c_{\bf {k}^\prime}\rangle ,\\& & \nonumber\\
\langle c_{-\bf k }c_{\bf k}\rangle & = & \frac{1}{2}\frac{\Delta_{\bf k}}{E_{\bf k}}\tanh(\beta E_{\bf k}/2).\nonumber
\end{eqnarray}
Setting $g_{\bf k} = 0$, Eq. (\ref{Ddconk}) reduces to the conventional BCS constraint that determines $\Delta_{\bf k}$.   However, it is shown below that $\Delta_{\bf k}$ cannot be determined from Eq. (\ref{Ddconk}) because $\langle \delta_{\bf k}\rangle$ and $\Delta_{\bf k}$ are in phase and increase simultaneously for an interaction with diagonal matrix elements.  Since $ 0 \leq \langle c_{-\bf k }c_{\bf k}\rangle \leq 1/2$, it is evident from Eq. (\ref{Ddconk}) that a large gap solution has a corresponding large deviation.

Using $H$ in Eq. (\ref{H}) gives the model expressions for the thermodynamic functions defined in Appendix B.  The thermodynamic potential $\Omega$ and the average internal energy ${\cal U} = \langle H\rangle$ are
\begin{eqnarray}\label{Omegak}
\Omega  & = & \Omega_{mf} + \langle H_d\rangle,\\& & \nonumber\\
\Omega_{mf} & = & -\sum_{\bf k}\{\frac{2}{\beta}\ln [2\cosh (\beta E_{\bf k} /2)] - \varepsilon_{\bf k} - \Delta_{\bf k} b_{\bf k}^{*}\},\nonumber
\end{eqnarray}
and
\begin{equation}\label{Uk}
{\cal U}  =  -\sum_{\bf k}[E_{\bf k }\tanh(\beta E_{\bf k }/2) - \varepsilon_{\bf k} - \Delta_{\bf k}b_{\bf k}^*] + \langle H_d\rangle.
\end{equation}
It should be noted that since the evaluation of $\langle H_d \rangle$ in Eq. (\ref{H}) is exact, the ${\cal U}(T)$ is exact for all $T$.  Although $\Omega(T)$ is approximate, it is expected to be accurate for $T \ll T_C$ since $\Omega(0) = {\cal U}(0)$.   The third thermodynamic function of interest is the entropy $S(T)$.  Inserting ${\cal U}$ and $\Omega$ from Eq. (\ref{Omegak}) into Eq. (\ref{entropy}) gives
\begin{eqnarray}\label{Sk}
S & = & \frac{1}{T}\sum_{\bf k}\{(2/\beta )\ln [2\cosh (\beta E_{\bf k}/2)] - \nonumber\\
 & & \nonumber\\
   & &\quad \quad E_{\bf k }\tanh(\beta E_{\bf k }/2)\}.
\end{eqnarray}
The standard Fermion gas form for $S$ applied in LTS,\cite{tinkham} is obtained using the identity $x = \ln[f(-x)/f(x)]$.  The error inherent in $\Omega$ accounts for the absence of the direct effect of $\langle H_d\rangle$ in $S$.  Such dependence is indirect, via $\Delta_{\bf k}$ determined self-consistently from the fixed point of $\Omega$.

\textbf{Random Phase Approximation:}
To resolve the system for non-negligible static fluctuation, we apply a random phase approximation(RPA) to $\langle H_d\rangle$ in Eq. (\ref{H}).  Assume the phases of the complex fluctuation components $\langle d_{\bf k}\rangle$ are random.   The only contribution to $\langle H_d\rangle$ in an ensemble average over the phases is from the phase independent ${\bf k} = {\bf k}'$ terms.  The fluctuation terms in $\langle H_d\rangle$ due to the off-diagonal elements of $\Gamma_{{\bf kk}'}$ are zero.  The average RPA value of the complex $\langle\delta_{\bf k}\rangle$ is also zero, but $|\langle\delta_{\bf k}\rangle| \neq 0$.  Noting that $\langle d_{\bf k}^{\dagger}\rangle = \langle d_{\bf k}\rangle^*$, the phase average RPA expressions are
\begin{eqnarray}
\langle H_d\rangle & = & \sum_{\bf k}\Gamma_{\bf k k}\left[|\langle d_{\bf k}\rangle|^2 + \frac{1}{4}\sigma_{\bf k}^2\right],\label{Hdrpa}\\& & \nonumber\\
|\langle\delta_{\bf k}\rangle|^2 & = & \sum_{{\bf k}^{\prime}}|\Gamma_{\bf k \bf {k}^\prime}\langle d_{{\bf k}^{\prime}}\rangle|^2.\label{deltarpa}
\end{eqnarray}
It should be noted that $\langle H_d\rangle$ can make a significant contribution to $H$ if $\Gamma_{\bf kk} \neq 0$, even when the $|\langle d_{\bf k}\rangle|^2$ terms are negligible.

\textbf{Interaction Matrix Structure:}  It is the goal here to formulate a minimal structure for $\bf \Gamma  = V + U$ which reflects the essential physics of cuprates.   The ${\bf U} = [U_{\bf k \bf {k}^\prime}]$ is formulated in Appendix A from a generic spin-singlet exchange interaction, with the effective doping dependent exchange $U(x) \propto J(x)$ given in Section II.

Since the doping range of the SC state of most cuprates is relatively small $(0.05 \lessapprox x \lessapprox 0.27)$, it is expected that a phonon mediated interaction $V$ has a significant contribution due to the formation of tight binding small polarons.  Small polaron formation in cuprates has three important consequences:\cite{callaway}  1) The lattice deformation in response to the charge variation tracks the electron hopping between Cu3d and O2p orbitals. 2) There is exponential reduction in the electronic bandwidth. 3) Electron hopping requires emission or absorption of phonons.  In the same order, relevance to our model is three fold: 1) It is assumed that the symmetry of $V$ is the same Cu3d-O2p bond symmetry as that of the exchange interaction.  2) There is a cut-off for the kinetic energy. 3) The diagonal elements $V_{\bf k k} = 0$.

An additional consideration is the repulsive electron-electron interaction $V_c$ included in $V = V_p + V_c$.  Since $V_p < 0$ and $V_c > 0$, the net phonon interaction $V(x) = -[|V_p(x)| - V_c(x)]$.  As the material changes with doping from an insulator to a metal, the $|V_p(x)|$ is expected to decrease with increasing $x$ as the strong small polaron interaction changes to a weaker Fermi-fluid-phonon form.\cite{zhou}  The $V_c(x)$ is also expected to decrease with increasing $x$ as screening increases.  Since $x$ dependence of the difference $|V_p(x)| - V_c(x)$ is reduced, $V$ is approximated by a constant.  Although $V$ involves both CuO and reservoir planes, 3D effects\cite{suominen} are incorporated here only in the strength of $V$.

In accordance with the above discussion, the interaction matrix elements are given by Eq. (\ref{Gamma1}), which is
\begin{eqnarray}\label{Gamma}
\Gamma_{\bf k \bf {k}^{\prime}} & = & - \Gamma_0(x)(1-\delta_{{\bf kk}'})\psi_{\bf k}\psi_{\bf {k}^{\prime}}^* - 2U_0(x)\delta_{\bf{kk}'},\nonumber\\& &\\
\Gamma_0(x) & = &  V_0 +  U_0(x)\quad U_0(x) = 4J(x).\nonumber
\end{eqnarray}
Gap symmetry is corroborated by numerous experiments on cuprates indicating a mixed $s$- and $d_{x^2 -y^2}$-wave gap,\cite{lee,keller,tsuei} and by general gauge and time-reversal symmetry breaking arguments.\cite{tsuei}  As shown after Eq. (\ref{psi2}) the symmetry factor $\psi_{\bf k}$ can be d-wave, or s-wave, or a complex linear combination.  All three choices have the same effect on the thermodynamic functions, which depend on $|\psi_{\bf k}|^2$.  However, a d-wave gap $\Delta_{\bf k} = \Delta\psi_{\bf k}^d$, with $\psi_{\bf k}^d = \cos(k_x) - \cos(k_y)$ enforces the Hubbard double occupancy restriction $\langle c_{{\bf r}\sigma}^{\dagger}c_{{\bf r}\sigma'}^{\dagger} \rangle  = 0$, which follows from the expression for $\langle c_{-\bf k}c_{\bf k}\rangle $ in Eq. (\ref{Ddconk}).

Using Eq. (\ref{Gamma}) in the definitions in Eq. (\ref{Deltak}) gives
\begin{eqnarray}\label{Ddk}
\Delta_{\bf k}  & \approx & \Delta\psi_{\bf k},\quad \Delta =  \Gamma_0\sum_{\bf k}\psi_{\bf k}^* b_{\bf k}\nonumber\\
& & \\
\langle\delta_{\bf k}\rangle & \approx & \delta\psi_{\bf k},\quad \delta = \Gamma_0\sum_{\bf k}\psi_{\bf k}^*\langle d_{\bf k}\rangle.\nonumber
\end{eqnarray}
The approximate $\Delta_{\bf k}$ and $\delta_{\bf k}$ assume that the sum of terms in $\Delta$ and $\delta$ is much larger than the single term arising from the diagonal elements of $\Gamma$.  Using the approximate forms of $\Delta_{\bf k}$ and $\delta_{\bf k}$, in the constraint Eq. (\ref{Ddconk}) gives
\begin{equation}\label{Ddconk1}
\pm\frac{|\delta|}{|\Delta|} = g = 1  - \frac{\Gamma_0}{2}\sum_{\bf k}\frac{|\psi_{\bf k}|^2}{E_{\bf k}}\tanh(\beta E_{\bf k}/2).
\end{equation}
Since $g$ is real, only magnitudes $|\delta|$ and $|\Delta|$ appear in the relative fluctuation ratio.  It is shown that $g \geq 0$ for an SC state to exist.

Using $\Delta_{\bf k}\approx \Delta\psi_{\bf k}$ from Eq. (\ref{Ddk}) to eliminate the $b_{\bf k}$ from $H_{mf}$ in Eq. (\ref{H}), and using Eq. (\ref{Gamma}) in the phase average RPA Eqs. (\ref{Hdrpa}) and (\ref{deltarpa}), the Hamiltonian assumes the form
\begin{eqnarray}\label{Hmod}
H & = &  H_{0} + \langle H_d\rangle,\\&{}\nonumber &\\
H_{0} & = & -\sum_{\bf k}[E_{\bf k }(1-n_{\bf k} - q_{\bf k}) - \varepsilon_{\bf k}] + \frac{|\Delta |^2}{\Gamma_0},\nonumber\\& &\nonumber \\
\langle H_d\rangle & = & - 2(1 + \alpha) U_0\frac{|\delta|^2}{\Gamma_0^2} - U_0\Sigma_1,\nonumber
\end{eqnarray}
where
\begin{eqnarray}\label{Sigmas}
\Sigma_1 & = & \frac{1}{2}\sum_{\bf k}\sigma_{\bf k}^2,\quad \alpha  =  \frac{\Sigma_{d2}}{\Sigma_{d1}} \ll 1,\quad\frac{|\delta|^2}{\Gamma_0^2} = \Sigma_{d1}\nonumber\\ & & \nonumber\\
\Sigma_{d1} & = & \sum_{\bf k}|\psi_{\bf k}|^2|\langle d_{\bf k}\rangle|^2, \quad \Sigma_{d2}  =  \sum_{\bf k}(1 - |\psi_{\bf k}|^2)|\langle d_{\bf k}\rangle|^2.\nonumber
\end{eqnarray}

The remainder of this article is based on the Hamiltonian in Eq. (\ref{Hmod}), coupled to the constraint Eq. (\ref{Ddconk1}).  The general symmetry dependent problem is outlined in Appendix D, where it is shown that the model not only produces small $U_0$ HTS solutions with the $|\delta|^2$ term neglected, but also extreme HTS solutions exist for weak interactions in the large fluctuation limit.  It is also concluded that the symmetry factor $\psi_{\bf k}$ does not fundamentally change the HTS thermodynamic properties which depend only on $|\psi_{\bf k}|^2$.

\section{AVERAGED SYMMETRY ANALYSIS}
In accordance with the above discussion, the sums are now transformed to integrals, with $|\psi_{\bf k}^2|$ replaced by the average $\langle|\psi_{\bf k}^2|\rangle_{av} = 1$, which sets $\alpha = 0$.  The interactions $V$ and $U$ may have different energy scales.  However, underdoped cuprates are narrow band insulators, with further narrowing due to the formation of small polarons.  Thus electronic kinetic energies are limited to values much less than the polaron cut-off at $T = 0$.  In the overdoped domain the band widths are larger but the effective pairing interaction is reduced, resulting in a small value of $T_C$.   Since the use of multiple cut-offs would not essentially change the HTS results over the relatively narrow doping dependent range of the SC state,  we invoke a kinetic energy cut-off.  This simplifying assumption is implemented by replacing electron energies $\varepsilon_{\bf k}$ by their $\bf k$-space angular averages $\langle\varepsilon_{\bf k}\rangle$, which are then bandwidth limited by $\langle\varepsilon_{\bf k}\rangle \leq \varepsilon_m = k_B T_m$.

In this section we focus on the resolution of the system when the effect of the fluctuation term with $|\delta|^2$ in Eq. (\ref{Hmod}) can be neglected.  However, since the gap $|\Delta|$ is determined from the thermodynamic potential fixed point equation, $\partial \Omega /\partial |\Delta| = 0$, the validity of the small fluctuation solution can only be assessed by retaining the $|\delta|^2$ in $\Omega$, and then neglecting it in the fixed point equation.

The integral forms of the thermodynamic potential $\Omega$ in Eq. (\ref{Omegak}), the internal energy ${\cal U}$ in Eq. (\ref{Uk}), with $\langle H_d\rangle$ from Eq. (\ref{Hmod}), and the entropy $S$ in Eq. (\ref{Sk}) are
\begin{eqnarray}
a\Omega (t, \phi ) & = & 1 - 4tI_0(t, \phi) + \frac{\phi^2}{\gamma} + a\langle H_d\rangle, \label{Omega} \\& &\nonumber \\
a{\cal U}(t, \phi) & = & 1 - 2I_3(t, \phi) + \frac{\phi^2}{\gamma} + a\langle H_d\rangle,\label{calU} \\& &\nonumber \\
a\langle H_d\rangle & = & - 2\nu\left(\frac{\phi g}{\gamma}\right)^2 - \frac{2\nu}{\chi}I_1(t, \phi), \nonumber\\ & & \nonumber\\
S(t, \phi) & = & 2k_B \chi[2I_0(t, \phi) - (1/t)I_3(t, \phi)].\label{S}
\end{eqnarray}
The constraint Eq. (\ref{Ddconk1}) assumes the form
\begin{equation}\label{gcon}
\pm\frac{|\delta|}{|\Delta|} = g(t, \phi) = 1 - \gamma I(t, \phi). 			
\end{equation}
The integrals $I_n(t, \phi$) are defined in Appendix E.  The scaled temperature and gap are
\begin{equation}\label{tphi}
t = \frac{T}{T_m} ,~~~ \phi (t) = \frac{|\Delta (t)|}{\varepsilon_m},
\end{equation}
and the material parameters are
\begin{eqnarray}\label{parameters}
\gamma & = & \eta  + \nu,\quad \eta  = N_0V_0, \quad \nu  =  N_0U_0 = 4N_0J \nonumber\\& &\\
\chi & = &  N_0 \varepsilon_m, \quad a = 1/(\chi \varepsilon_m). \nonumber
\end{eqnarray}
The $N_0$ is the average density of electron states in the energy integration interval $[-\varepsilon_m , \varepsilon_m]$.  The $\eta = \eta_p -\eta_c$ is the effective electron-phonon interaction parameter $\eta_p$ reduced by the repulsive coulomb parameter $\eta_c$.  As discussed before Eq. (\ref{Gamma}), there is at least partial canceling of the $x$ dependence of $\eta$.   It is possible that other parameters, cut-off energy $\varepsilon_m$, density of states $N_0$, and chemical potential $\mu$ are also functions of $x$.  However, lacking explicit information, the model is kept reasonably simple by neglecting doping dependence of the set $[\eta, \varepsilon_m, N_0, \mu]$.  Excellent agreement with experiment in Section V confirms the validity of implementing this approximation.
 							
\textbf{Gap Equation:} The thermodynamic potential (\ref{Omega}) contains two unknowns, $|\Delta (t)|$ and $|\delta(t)|$.   The constraints $g = 1 - \gamma I$, and $\partial\Omega/ \partial\phi^2 |_t = 0 $ determine $g$ and $\phi$, self-consistently, as functions of the parameters $[\eta, \nu, \chi]$ at each value of $t$.  From Eq. (\ref{Omega}) one obtains the fixed point equation
\begin{equation}\label{dOmega}
a\frac{\partial\Omega}{\partial\phi^2}|_t = \frac{g}{\gamma} - 2\nu\left[\left(\frac{g}{\gamma}\right)^2 + I_4\frac{g}{\gamma}\right] - \frac{\nu}{\chi}I_2 = 0,
\end{equation}
where the integrals $I_2$ and $I_4$ are defined in Appendix E.  The first term is the mean field term, the second and third terms are from the $|\delta|^2$ fluctuation term, and the last term is due to the constant diagonal elements of the effective exchange interaction $U$.  In the limit $\nu = 0$, Eq. (\ref{dOmega}) gives $g = 1 - \eta I(t,\phi) = 0$, which is the LTS BCS constraint for non-zero $t$.

Neglecting the fluctuation terms, Eq. (\ref{dOmega}) reduces to
\begin{equation}\label{keysm}
\frac{g}{\gamma} = \frac{1}{\gamma} - I(t,\phi) = \frac{\nu}{\chi}I_2(t, \phi) \geq 0,
\end{equation}
which is valid for parameters satisfying
\begin{equation}\label{nusm}
2\nu\left[\frac{\nu}{\chi}I_2(t, \phi) + I_4(t, \phi)\right] \ll 1.
\end{equation}
Solutions of Eq. (\ref{keysm}) exist for $\gamma = \eta + \nu > 0$, since the integrals $I \geq 0$ and $I_2 \geq 0$.  Thus the exchange interaction $\nu$ is so effective that even if $\eta = \eta_p -\eta_c < 0$, i.e. when the coulomb interaction overrides the phonon interaction, a SC solution exists for values of $\nu$ as long as $\eta + \nu > 0$.

It is elucidating to consider approximate implicit solutions of Eq. (\ref{keysm}) for the interaction limit $\gamma /[1 - g(t)] < 0.3$ at $t = 0$ and $t = t_C$.  Using Eqs. (\ref{Itzero}) and (\ref{Iphi0}) gives
\begin{equation}\label{expsol}
\phi (0) = 2e^{-1/\gamma}e^{g(0)/\gamma},~~~t_C = 1.134e^{-1/\gamma}e^{g(t_C)/\gamma}.
\end{equation}
The BCS MF results for $\Delta (0)$ and $T_C$ are obtained for $g(\nu = 0) = 0$.  As $\nu$ increases from zero, the $g(t) > 0$ initially results in an exponential $g/\gamma$ increase in $\Delta (0)$ and $T_C$.  For larger $\nu$ the dependence of $\Delta(t)$ and $T_C$ on $\nu$ is linear over a relatively broad range of $\nu$ values.

The linearity is found analytically under certain conditions.   It follows from Eq. (\ref{keysm}), using Eqs. (\ref{Itzero}) and Eq. (\ref{Iphizero}) that
\begin{eqnarray}\label{ggam}
\frac{g(0, \phi)}{\gamma} & = & \frac{1-\pi/4}{\chi}\frac{\nu}{\phi},\quad \mbox{for}\quad  \phi(0) \ll 1\nonumber\\& &\\
\frac{g(t, 0)}{\gamma} & = & \frac{C_1(t)}{\chi}\frac{\nu}{t},\quad \mbox{for}\quad  t \approx t_C \ll 1.\nonumber
\end{eqnarray}
where $0.0738 < C_1(t) \lessapprox 0.2$.  It is worth noting that $\varepsilon_m$ and $N_0$ cancel in both expressions in Eq. (\ref{ggam}); hence the enhancement of the gap and $t_C$, relative to the BCS MF values, is due only to the exchange interaction $\propto U_0 > 0$, independent of the cutoff energy and the average density of states.  Using $g(0)$ and $g(t_C)$ from Eq. (\ref{ggam}) in Eq. (\ref{expsol}), leads to approximate linear relations for the ratio of the $t = 0$ gap and $t_C$  at $\nu$ and $\nu_o$.  They are
\begin{equation}\label{linear}
\frac{\phi(\nu)}{\phi(\nu_o)} = \frac{t_C(\nu)}{t_C(\nu_o)} = \frac{\nu}{\nu_o},
\end{equation}
which are independent of $N_0$ and $T_m$.  The importance of this positive linearity to HTS, in stark contrast to the negative exponential dependence in LTS, cannot be over emphasized.  It was shown in Section II to be an essential characteristic for the doping dependence of the gap $|\Delta(x)|$, and $T_C(x)$ observed in cuprates.  It is also found numerically that Eq. (\ref{linear}) holds over a relatively broad range of $\nu$, including values that give significant fluctuations requiring the solution of Eq. (\ref{dOmega}).  Using Eq. (\ref{JProbSC}), and setting $\nu_o = \nu(x_{op} = x_1 + w/2) = \nu_m$, the doping dependence of the effective exchange energy is
\begin{equation}\label{nux}
\displaystyle{\frac{\nu(x)}{\nu_m}} = 4\Pi_h(x)\Pi_e(x).
\end{equation}
Thus $|\Delta(x)|$, and $T_C(x)$ are linked to $x$.

\textbf{Critical Field, and Specific Heat:} The condensation energy
\begin{equation}\label{condens}
\Delta\Omega(t, \phi) = \Omega(t, \phi) - \Omega(t, 0),
\end{equation}
defines a thermodynamic critical magnetic field $H_C$ by
\begin{equation}\label{Hcdef}
(1/2)\mu_o H_C^2(t) = |\Delta\Omega(t, \phi)|.
\end{equation}
At low temperatures, $T \ll T_C$, for $\nu > 0$ Eq. (\ref{condens2}) yields the linear temperature dependence
\begin{eqnarray}\label{Hc}
\frac{H_C(T)}{H_C(0)} & = & 1 - B(\phi)\frac{T}{T_C},\\& &\nonumber\\
B(\phi) & = & \frac{\nu}{\chi}\frac{\ln 4 - 1}{|a\Omega(0, \phi)|}\frac{T_C}{T_m}.\nonumber
\end{eqnarray}
It is shown in section V that for $\nu > 10^{-3}$ the slope $B(\phi)$ lies in the range $[0.4, 0.6]$, in agreement with cuprate HTS observation.  The $\nu = 0$ LTS limit in Eq. (\ref{condens2}) gives a very different $H_C(T)/H_C(0) = 1 - 1.06(T/T_C)^2$.

The entropy determines the specific heat $C = T\partial S/\partial T$ for $\mu$ constant.  Using the integral relations (\ref{intrelate}) in Eq. (\ref{S}) gives the form
\begin{equation}\label{C1}
C(t) = \frac{k_B\chi}{t^2}\int_{0}^{1}dy\left[Y^2 -\frac{t}{2}\frac{\partial\phi^2}{\partial t}\right]\cosh^{-2}\left(\frac{Y}{2t}\right).
\end{equation}
The normal state specific heat $C_n(t) = C(t,0)$ is given by the first term in Eq. (\ref{C1}) with $Y = y$.  The discontinuity $\Delta C(t_C) = C(t_C, \phi \rightarrow 0 - C_n(t_C)$ at $t_C$, obtained from Eq. (\ref{C1}) is
\begin{equation}\label{DC}
\Delta C(t_C) = -k_B\chi\tanh\left(\frac{1}{2t_C}\right)\frac{\partial\phi^2}{\partial t}|_{t_C}
\end{equation}
For $\nu > 0$ there is a large increase in the discontinuity due to the large slope $\partial \phi^2 /\partial t|_{t_C} < 0$.  A quantitative comparison between the HTS and LTS $C(t)$ is given in section V.  General relationships between $C$ and other forms of specific heat are developed in Appendix C.
\section{NUMERICAL SOLUTIONS AND COMPARISON WITH EXPERIMENT}

\begin{figure}
\includegraphics[width = 3.2in]{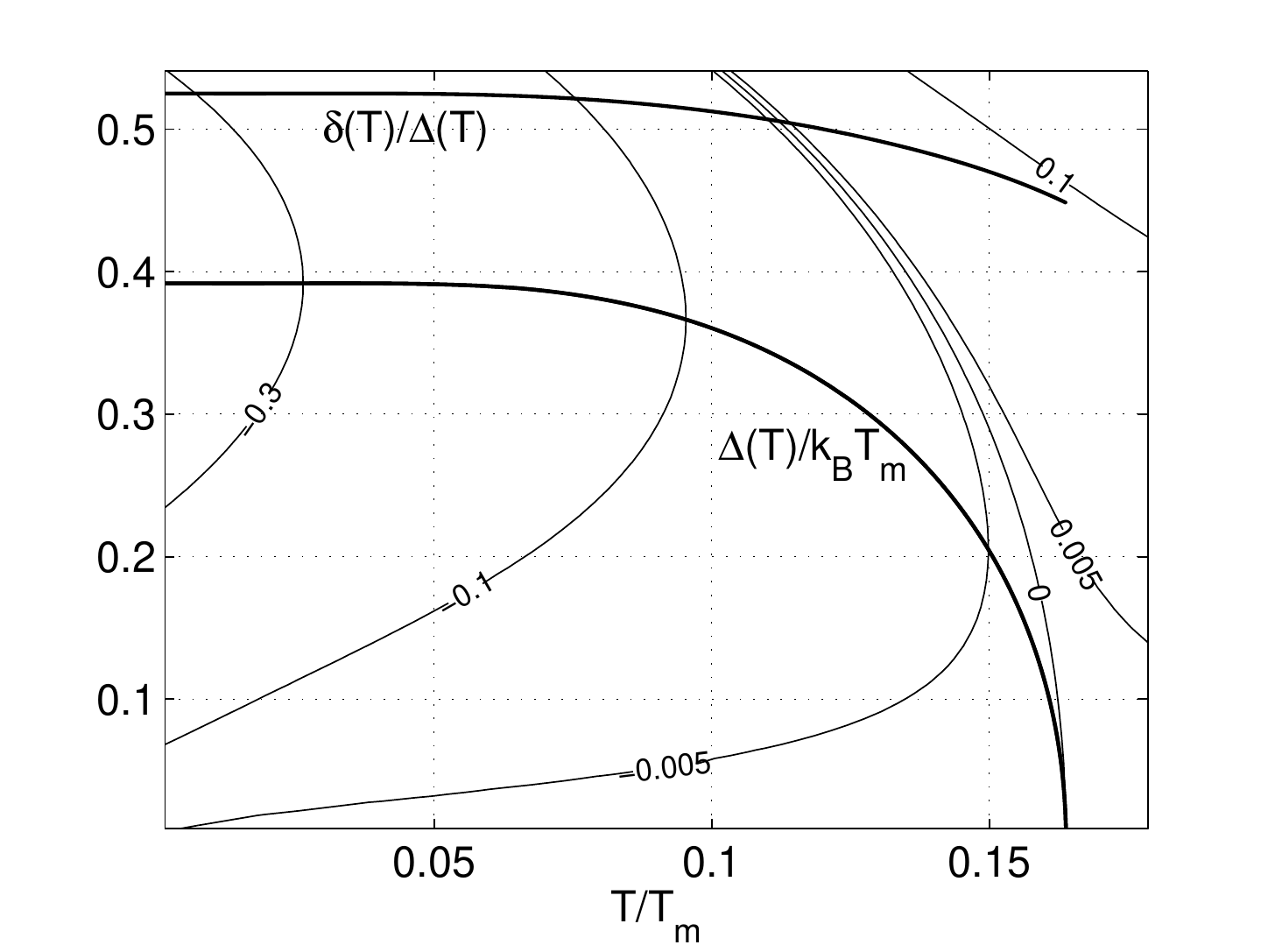}
\caption{\label{Fig2} SOPT solutions for $\phi(t)$ and $g(t) = |\delta(t)/\Delta(t)|$ of Eq. (\ref{gkey}) plotted for $\eta = 0.25, \nu = 0.035$, and $\chi = 0.01$.  The contours are the scaled condensation energy.}
\end{figure}

\begin{figure}
\includegraphics[width = 3.2in]{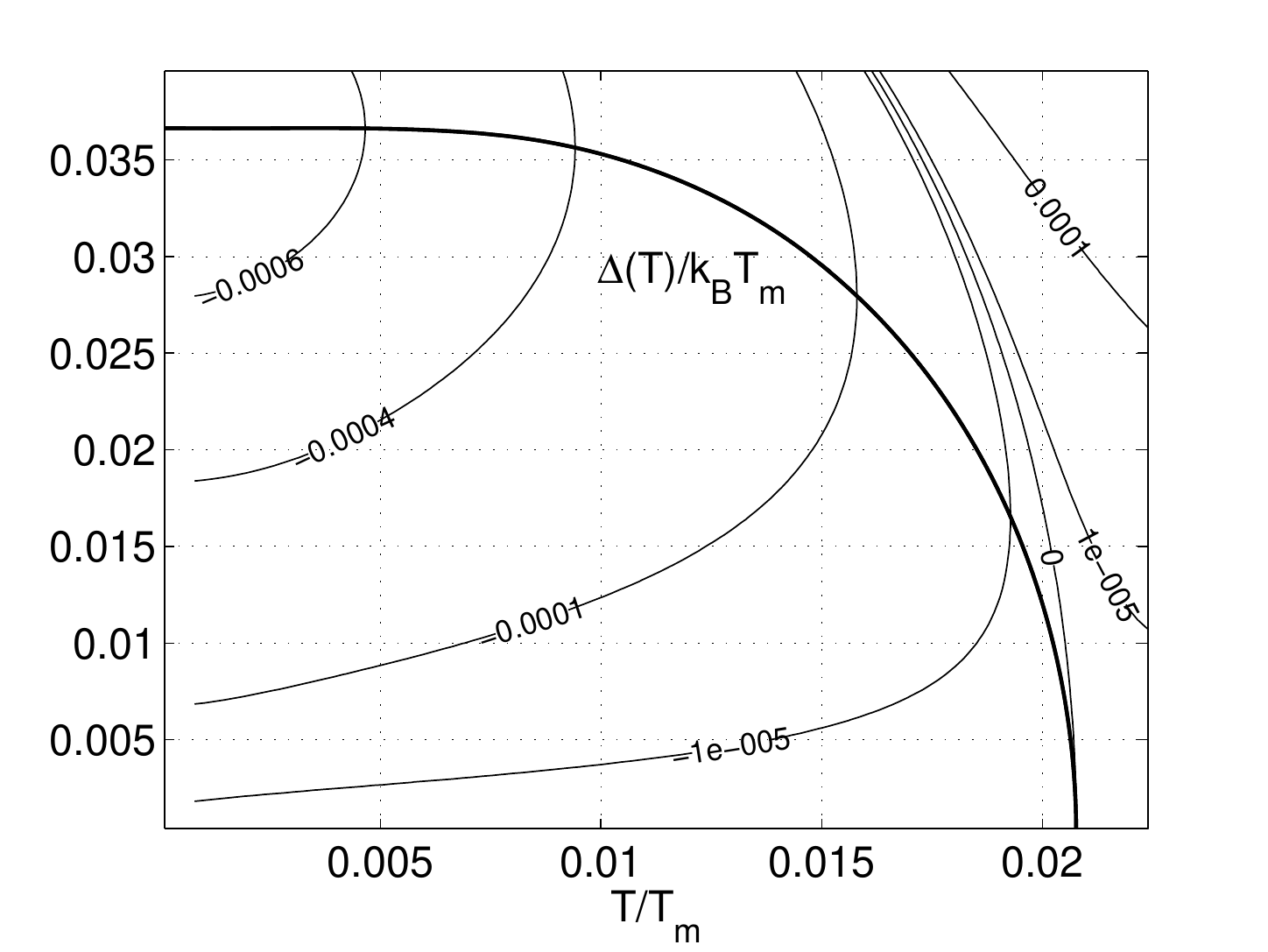}
\caption{\label{Fig3} LTS mean field solutions plotted for $\eta = 0.25, \nu = 0.0$, and $\chi = 0.01$.}
\end{figure}

The numerical solution of Eq. (\ref{keysm}) is facilitated by starting at t = 0, using the exact integrals in Eq. (\ref{Itzero}) and then increasing $t$.  The minimum free energy SC gap $\Delta (t)$ is shown in Fig. \ref{Fig2}.  Contours are the condensation energy $\Delta \Omega (t)$ in Eq. (\ref{condens}) divided by $\chi \varepsilon_m$.  Contours to the left(right) of the zero contour are negative(positive).  Fig. \ref{Fig2} is plotted for a cutoff parameter $\chi = 0.01$, phonon parameter $\eta = 0.25$, and a comparatively small exchange parameter $\nu = 0.035 = 0.14\eta$.  The significant points are $\phi(0) = \Delta(0)/k_BT_m = 0.392$,  $t_C = T_C/T_m = 0.164$, and gap ratio $\Delta(0)/k_B T_C = 2.39$.  For comparison, using the same values of $\chi$ and $\eta$, Fig. \ref{Fig3} is the $\nu = 0.0$ LTS MF solution, with $\phi_{mf}(0) = 0.0366$,  $t_{Cmf} = 0.021$, and $\Delta (0)/(k_B T_C) = 1.76$.  The enhancements relative to the LTS MF values are $\Delta (0)/ \Delta_{mf}(0) = 10.7$ and $T_C /T_{Cmf} = 7.81$.  Thus a 12K LTS becomes a 94K HTS.  The remarkable effectiveness of the diagonal matrix elements of the exchange interaction is clearly evident by comparing these enhancements with the modest mean field only enhancement $\Delta_{mf}(T = 0,\eta +\nu)/\Delta_{mf}(T= 0, \eta) = T_{Cmf}(\eta + \nu)/T_{Cmf}(\eta) \approx 1.63$.  Large enhancement for all $t \leq t_C$ is concomitant with relatively large values of $g(t) = |\delta(t)/\Delta(t)|$, which varies from $g(0) = 0.52$ to $g(t_C) = 0.45$.

\begin{figure}
\includegraphics[width = 3.2in]{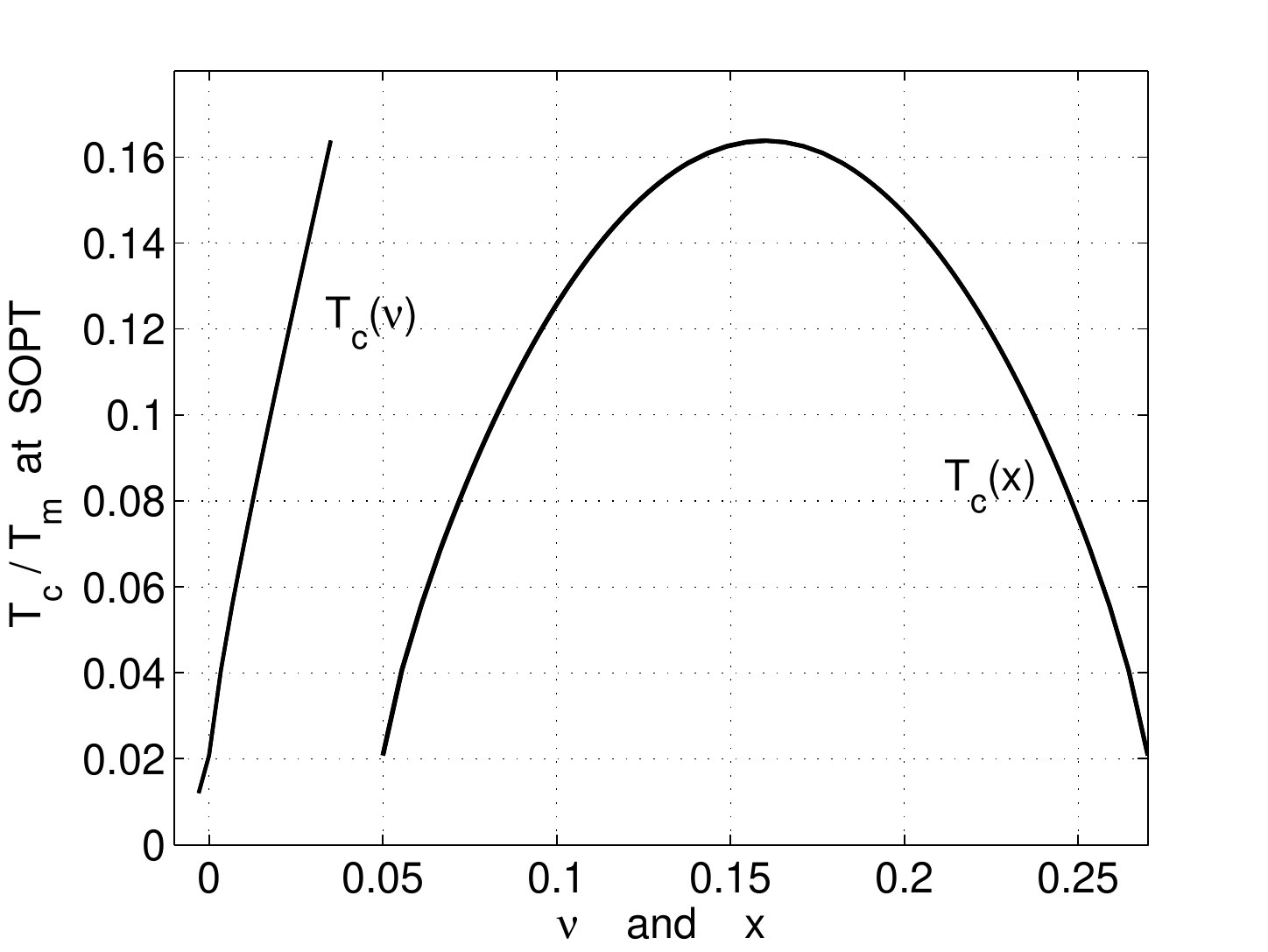}
\caption{\label{Fig4} The scaled transition temperature $t_c(\nu)$ for SOPT's is plotted as a function of the interaction parameter $\nu$, and using Eq.(\ref{nux}), as a phase boundary $t_c(x)$, with $x$ the hole doping concentration in the range $[0.05, 0.27]$.  The curve is for $\eta = 0.25$, $\chi = 0.01$, with $\nu_m = 0.035$.  The maximum is $T_C(x_{op})/T_m = 0.164$.}
\end{figure}

The SOPT $t_C$ is plotted in Fig. \ref{Fig4} as a function of $\nu$, and using Eq. (\ref{nux}) as a typical cuprate second order phase transition boundary $t_C(x)$ for hole doping. \cite{keller,lee,hufner}  An abrupt transition to the insulator(metal) state occurs for $x$ just outside the $[x_1, x_2]$ range.    The $t_C(x)$ plotted in the doping range $[0.05, 0.27]$ for $\eta = 0.25, \nu_m = 0.14\eta $, where $t_C(\nu)$ is a linear function of $\nu$, has the parabolic shape observed in rather broad collection of cuprates: Bi$_2$Sr$_2$CaCu$_2$O$_{8+\delta}$, YBa$_2$Cu$_3$O$_{7-\delta}$,  Tl$_2$Ba$_2$CuO$_{6+\delta}$, HgBa$_2$CuO$_{4+\delta}$ which have same $T_C(x_{op}) \sim 90-95$K, but with different numbers, $n = 1,2,3$, of Cu-layers per unit cell.\cite{hufner}  For this value of $w = x_2 - x_1 =  0.22$, the curve is often referred to as the empirical "universal curve".  A more appropriate designation is "universal parabolic".  The general form given by Eq. (\ref{universe}) and the equivalent Eq. (\ref{universal}), is a fundamental cuprate characteristic based on state occupation probability arguments.  Other cuprates exhibit a parabolic phase boundary, but with a different values for $[x_1, x_{op}, x_2]$, and maximum $T_C$.  For example. other measurements on the two-layer Bi$_2$Sr$_2$CaCu$_2$O$_{8+\delta}$ give $[x_1 \approx 0.06, x_{op} \approx 0.12]$,\cite{karppinen} but the shape of the SC phase boundary remains parabolic.  The phase boundaries of the single layer Bi$_2$Sr$_2$LaCuO$_{6+\delta}$ and Bi$_2$PbSr$_2$LaCuO${6+\delta}$ are also parabolic with $[x_1 \approx 0.11, x_2 \approx 0.235]$.\cite{schneider}

The lower $T_C(x)$ for these cuprates with $T_C(x_{op}) < 40^oK$ is obtained in the model by reducing the value of $\eta$ and/or $\nu$, and changing the range of $x$.  However, the scaled curves are independent of $\eta$ and are given by Eq. (\ref{universal}) below.  Doping dependence of $N_0(x)$\cite{schneider} and $\mu = \mu(x)$\cite{shen} may play a role in determining the SC phase transition range $[x_1, x_2]$, and they probably introduce some asymmetry into the $t_C(x)$ boundary.  Also, the sign change of $d\mu(x)/dx$ contributes to the observed asymmetry of the phase boundary for hole versus electron doping.

\begin{figure}
\includegraphics[width = 3.2in]{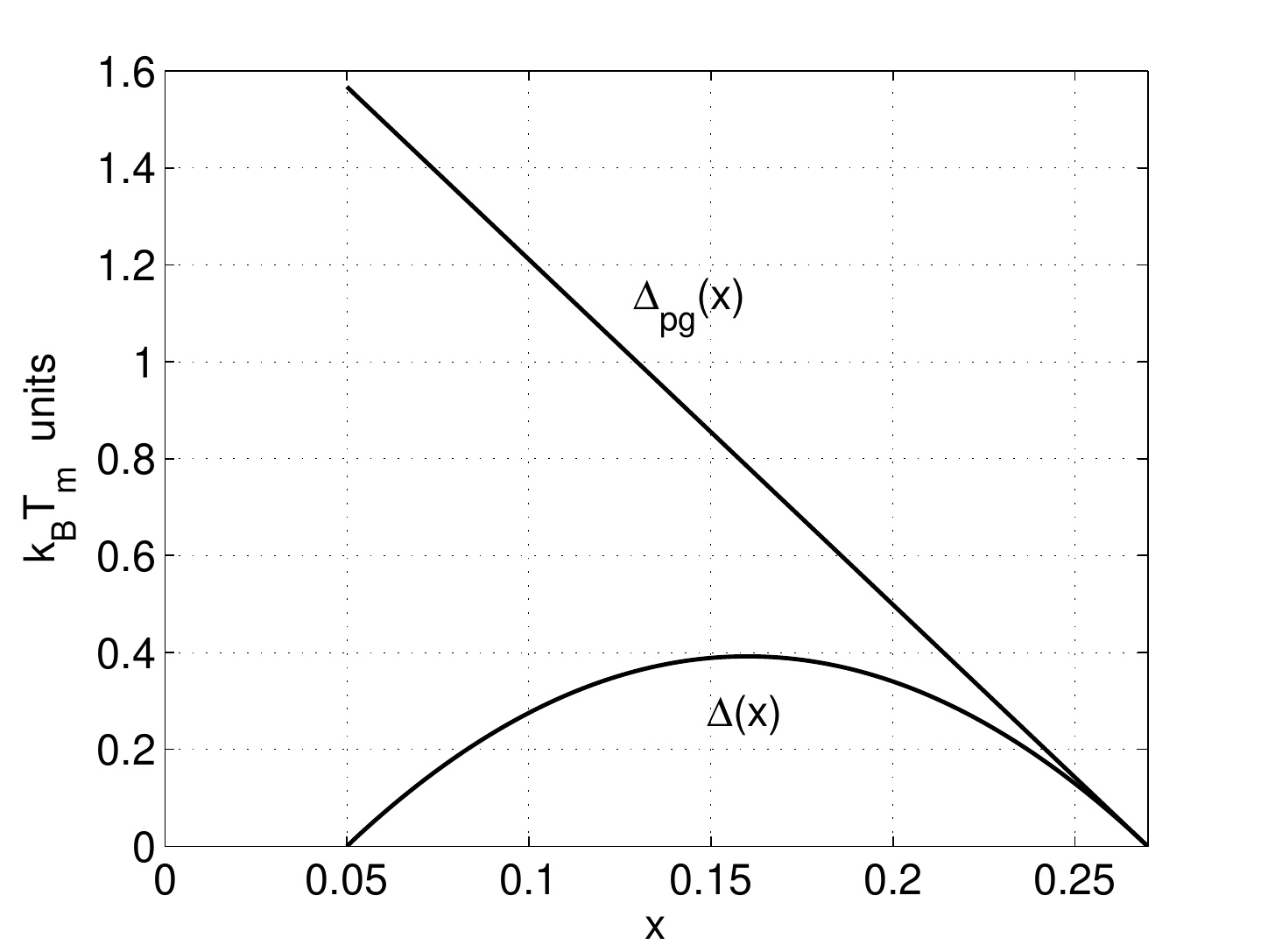}
\caption{\label{Fig5} The SC gap $\Delta(T = 0, x)$ and the pseudogap $\Delta_{pg}(x)$ are plotted in units of $k_BT_m$ as a function of the doping concentration $x$ for the same parameter values as in Fig. \ref{Fig2}.  The maximum gap is $\Delta(x_{op})/k_BT_m = 0.392$. }
\end{figure}

The $T = 0$ doping dependent SC gap $\Delta(x)$ and the pseudogap $\Delta_{pg}(x)$ are plotted in Fig. \ref{Fig5} for $\eta = 0.25, \nu_m = 0.140\eta$.  The gap $\Delta(x)$ has the same shape as $T_C(x)$, since both are linear functions of $\nu(x)$.  The $x$ dependence of both $\Delta(x)$ and $T_C(x)$ is given by Eq. (\ref{universe}), which can be written in the form

\begin{equation}\label{universal}
\frac{T_C(x)}{T_C(x_{op})} = \frac{\Delta(x, 0)}{\Delta(x_{op})} = 1 - \left[1-\frac{2}{w}(x - x_1)\right]^2,
\end{equation}
where $w = x_2 - x_1$.  For the case $x_1 = 0.05, w = 0.22$, Eq. (\ref{universal}) is identical to the empirical,\cite{presland,tallon2,hufner} doping dependent function observed in Bi2212, YB123, Tl2201, Hg1201, listed above, with $T_C(x_{op}) \sim 90-95$K, $\Delta_{exp}(x_{op}, 0) \approx 21 \pm 1$mev, and $\Delta_{exp}(0)/k_BT_C \approx 2.5\pm 0.15$, as summarized in Ref. \cite{hufner}.

The model $\Delta(0,x)$ curve fits the experimental $\Delta_{exp}(0, x)$ curve for a cutoff energy $\varepsilon_m = \Delta_{exp}(x_{op})/\phi(x_{op}) \approx 51$mev, at $x_{op} = 0.16 \approx 1/6$.  Using $26 \mbox{mev} \simeq 300$K gives $T_m \approx 589$K.  The $T_m$ found from Fig. \ref{Fig3} is $T_m = T_C(\mbox{exp},x_{op})/t_c(x_{op}) \approx 92.5/0.164 = 564$K.  The relative difference in $T_m$ is $4.3\%$, which is within the experimental error in $\Delta_{\mbox{exp}}(x_{op}, 0)$ and the spread of $T_C(\mbox{exp})$.  The effective density of states $N_0 = \chi/\varepsilon_m \approx 1/(5\mbox{ev})$, which is that of a metal, i.e. $N_0 \sim 1/\epsilon_F$.  This is consistent with BCS superconductor behavior at optimal doping.\cite{lee,hufner} Using the $N_0$ gives the exchange constant at optimal doping $J(x_{op}) = 0.25U_0(x_{op}) \approx 0.0438\mbox{ev}$, and $V_0 \approx 1.25 \mbox{ev}$.  Comparison with the undoped Cu3d-Cu3d exchange $J_{dd} = 0.13\mbox{ev}$ gives $J(x_{op})/J_{dd} \approx 1/3$.  Relating this result for the cuprates in Ref. \cite{hufner} with the commonly used t-J model ratio $J_{dd}/t \approx 1/3$,\cite{lee} gives the geometric mean
\begin{equation}\label{JddJpd}
J_{dd} = \sqrt{tJ(x_{op})},\quad \mbox{for}\quad x_{op} \approx 1/6.
\end{equation}
For these cuprates the competition between kinetic energy per unit area and exchange energy is characterized by $2x_{op}t \approx J_{dd}$.

The pseudogap given by Eq. (\ref{pg}) is
\begin{equation}\label{pg2}
\Delta_{pg}(x) = 4\Delta(x_{op})\left[\frac{x_2 - x}{w}\right].
\end{equation}
Using the above $\Delta(x_{op})$ gives $\Delta_{pg} = 42 \pm 4$mev, and  $\Delta_{pg}(x_1) = 4\Delta(x_{op}) = 84 \pm 4$ mev.  This is within experimental error of the measured pseudogap $\Delta_{pg} = 76 \pm 4$, extrapolated to $x_1 = 0.05$.\cite{hufner}  With this starting value, $\Delta_{pg}(x)$ fits the data in the SC range $0.05 < x < 0.27$.

\begin{figure}
\includegraphics[width = 3.2in]{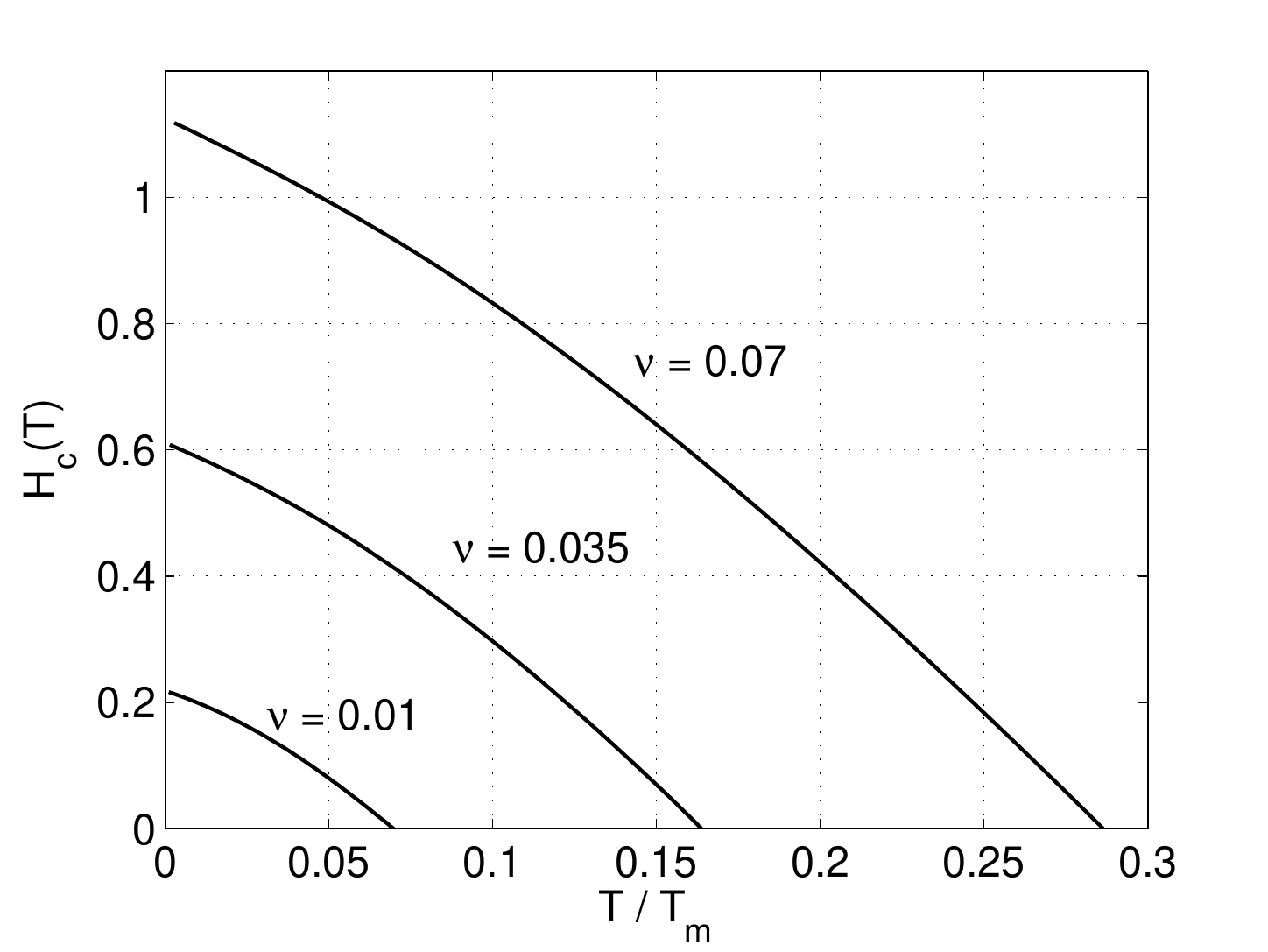}
\caption{\label{Fig6} Scaled thermodynamic critical fields $H_C(T)$ are shown as a function of $T/T_m$ for $\eta = 0.25, \chi = 0.01$ and three values of $\nu$.  The vertical scale is in units of $\chi\sqrt{2/(N_0\mu_0)}$}
\end{figure}
\begin{figure}
\includegraphics[width = 3.2in]{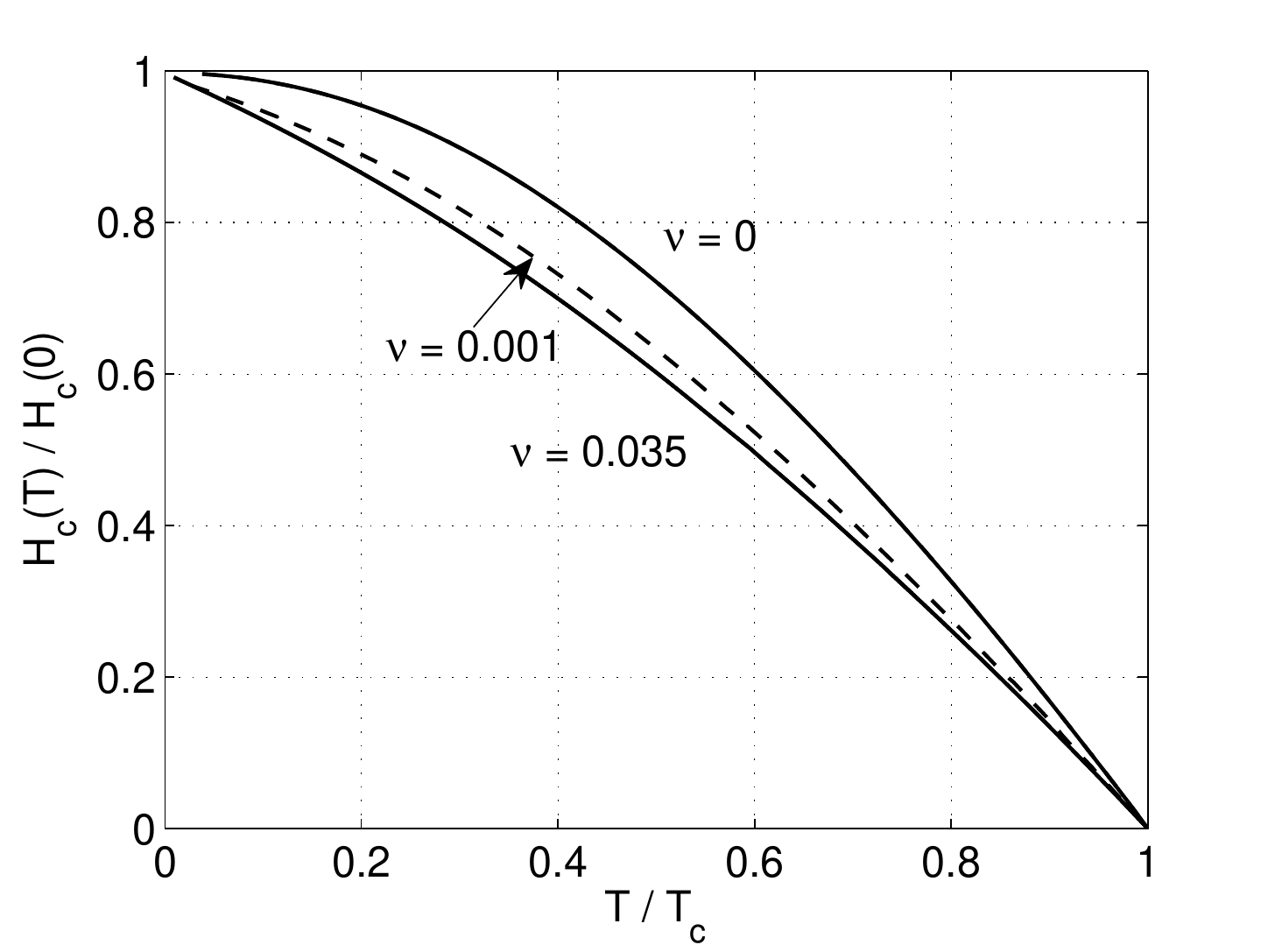}
\caption{\label{Fig7} Thermodynamic critical fields $H_C(T)/H_C(0)$ are shown as a function of $T/T_C$ for the same parameter values in Fig. \ref{Fig6}.  The $\nu = 0$ curve is the LTS critical field of the BCS model, which lies slightly below that of the two-fluid model $H_C(T)/H_C(0) = 1 - (T/T_C)^2$.}
\end{figure}

The thermodynamic critical field $H_C(T)$ is shown in Fig. \ref{Fig6} versus $T/T_m$ and in Fig. \ref{Fig7} versus $T/T_C$.  The curves in Fig. \ref{Fig6}, with $\nu = 0.07$, are based on the full Eq. (\ref{gkey}).  For $T \lessapprox T_C/3$ there is significant linear slope for all $\nu(x) > 0$, which has almost no dependence on $x$.  The field $H_C$ is related to the magnetic field penetration depth $\lambda(T)$ by a simple argument.  In cuprates the existence of vortices with effective penetration area $\pi \lambda^2$ are pierced by constant quantized flux for $T \ll T_C$.  Thus the flux $\pi\lambda^2(T)H_C(T)$ is constant, and Eq. (\ref{Hc}) gives the form
\begin{equation}\label{hclambda}
\frac{\lambda^2(0)}{\lambda^2 (T)} = 1 - B(\phi)\frac{T}{T_C}.
\end{equation}
For the range $10^{-3} \lessapprox \nu \lessapprox 0.035$, we obtain a slope $B \approx 0.4 - 0.6$.  This is in quantitative agreement with the observed \cite{boyce,hardy,jacobs,sflee,hosseini,stajic} linear $T$ dependence of $\lambda^{-2}(T)$ with slopes $B \approx 0.5 \pm 0.1$.

\begin{figure}
\includegraphics[width = 3.2in]{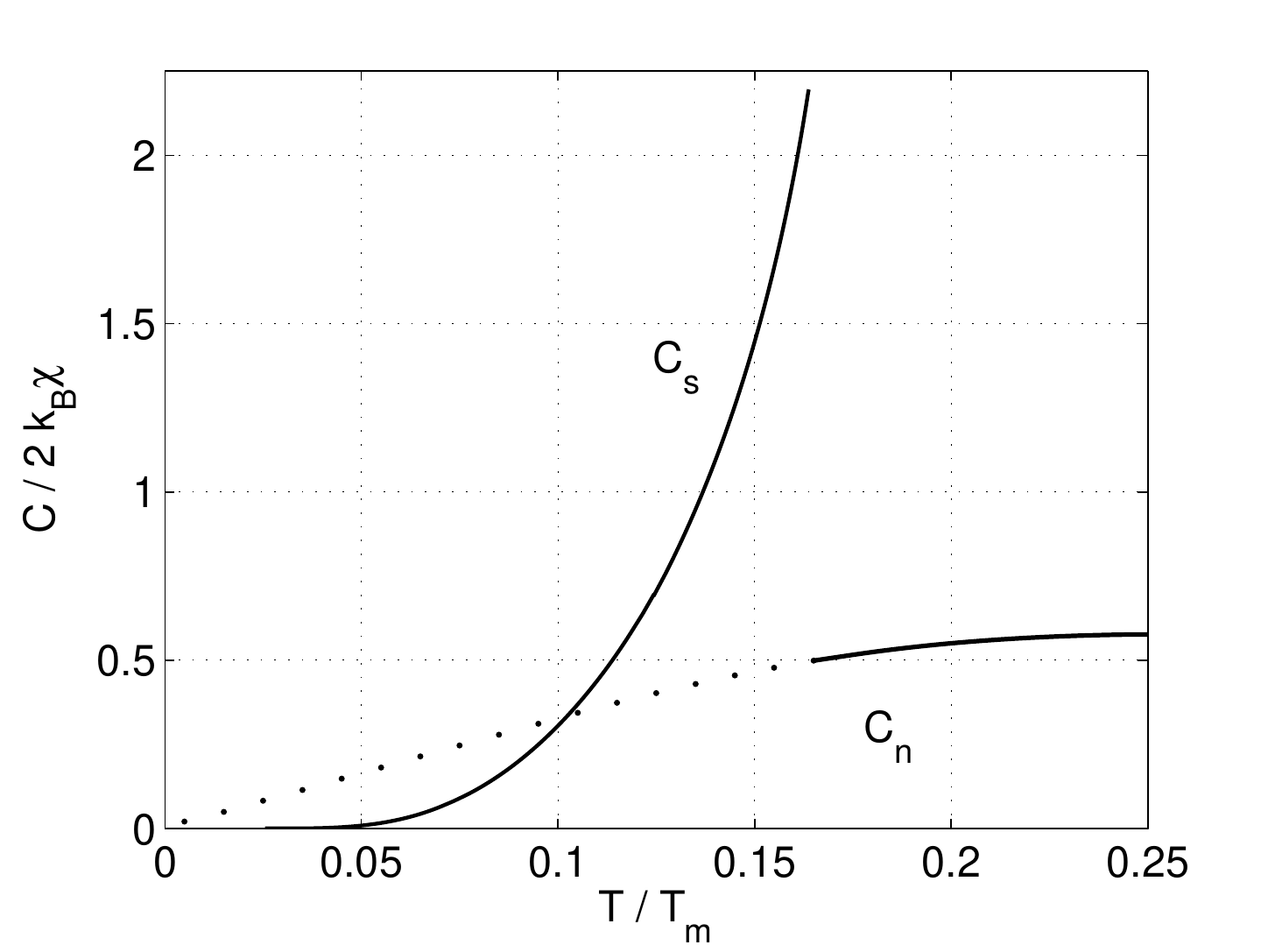}
\caption{\label{Fig8} HTS specific heat $C$, calculated from Eq. (\ref{S}), is shown as a function of $t$ for parameter values in Fig. \ref{Fig2}.  The discontinuity at $T_C$ is 2.36 times the corresponding LTS value.}
\end{figure}

In the SOPT domain the specific heat $C(t) = t\partial S/\partial t$ increases with increasing $\nu(x)$.  As a consequence the specific heat discontinuity at $t_C$ increases with hole doping, reaching a maximum at optimal doping, as observed.\cite{lee}   The normalized HTS specific heat $C_s(\nu = 0.035, \phi)/(2k_B\chi)$, and the normal state $C_n(\nu, \phi = 0)/(2k_B\chi)$ are plotted in Fig. \ref{Fig8}.  The discontinuity at $T_C$ is $(C_s - C_n)/C_n = 3.38$.  The corresponding discontinuity for the LTS specific heat $C(\nu = 0, \phi)$ is $1.43$.  Thus the model gives the observed, large comparative, difference between LTS and HTS specific heats.

\begin{figure}
\includegraphics[width = 3.2in]{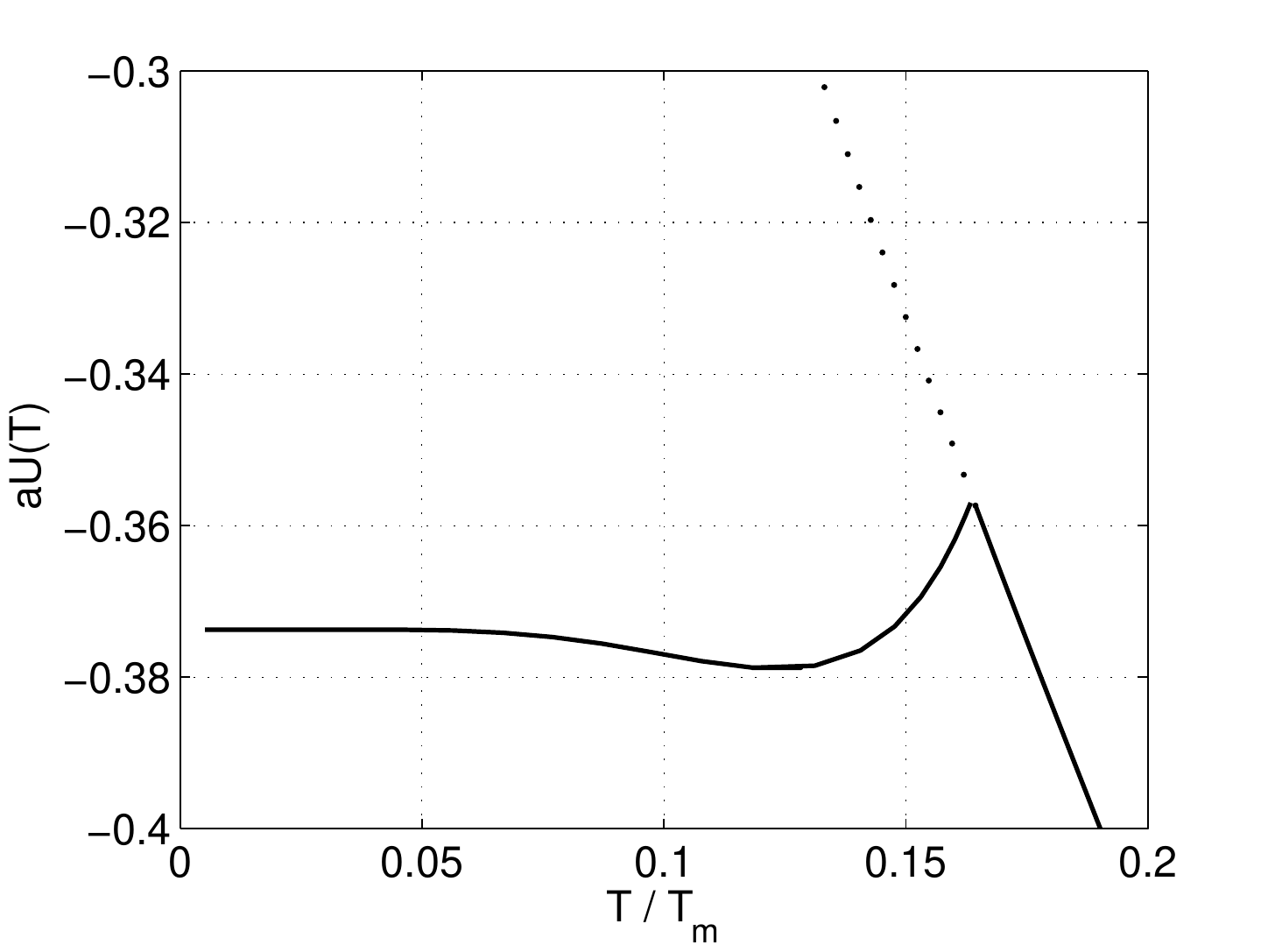}
\caption{\label{Fig9} The internal energy $\cal U$ from Eq. (\ref{calU}) is shown as a function of $t$ for parameter values in Fig. \ref{Fig2}}.
\end{figure}

The HTS internal energy ${\cal U}(T, \nu)$ shown in Fig. \ref{Fig9} for $\nu = 0.035$ exhibits anomalous behavior in contrast to the LTS internal energy ${\cal U}(T, 0)$, which exhibits a monotonic increase with $T$ for both the SC and normal states (See Fig. 3.3 in Tinkham\cite{tinkham}).  The unusual dip in ${\cal U}(T, \nu)$ in the range $t_1 \approx 0.07 < t < t_2 \approx 0.14 < t_C = 0.164$ indicates a temperature range $[T_1, T_2]$ of stronger effective exchange pairing than that for $T < T_1$.  Entering the dip from the left, energy is transferred to non-electronic parts of the system.   As $T \rightarrow T_C$ the system is absorbing energy as SC pairs are breaking.  In the normal state $T > T_C$ energy is again emitted, indicating some form of order stemming from the exchange parameter $\nu$.  Saturation, due to the kinetic energy cut-off, occurs for $T \gg T_C$ with ${\cal U}(T, \nu)\rightarrow 1 - \nu/\chi$.

The oxygen isotope effect on $T_C$ is characterized by the coefficient $\alpha_{iso}  = - \partial\ln t_C/\partial\ln M$, with $M$ the oxygen isotope mass.  Quantitative assessment of the effect may be complicated and involve the effective carrier mass $m^*$ on the overdoped side.\cite{alexandrov, keller} However, making the simple assumption that $dM \propto -dx$ to preserve charge neutrality during hole doping, it follows from $t_C(x)$ plots that $\alpha_{iso}$ is positive for $x < x_{op}$, negative for $x > x_{op}$, and independent of $x$ near $x_{op}$.  Thus, without rigorous derivation, the model qualitatively gives the observed doping dependent evolution of the isotope effect on $T_C$.

\section{LARGE FLUCTUATIONS}
Comparisons in Section V to cuprate HTS are based on Eq. (\ref{keysm}) in which fluctuations were neglected.  Here we consider solutions of Eq. (\ref{dOmega}) to determine the deviation from solutions of Eq. (\ref{keysm}) for the case considered in Section V with the maximum $\nu(x_{op}) = 0.035$,  and to show the effect of large fluctuations.   Solving Eq. (\ref{dOmega}) for $g$ leads to the key integral equation for $\phi(t)$.  It is
\begin{eqnarray}\label{gkey}
\frac{1}{\gamma} - I(t, \phi) & = & \frac{1}{2}h_{\nu} (t, \phi)\left[1 \pm \sqrt{1 - Q(t, \phi)}\right],\\& &\nonumber\\
Q(t, \phi) & = & \frac{2}{\chi}\frac{I_2(t, \phi)}{h_{\nu}^2(t,\phi)},\quad h_{\nu}(t, \phi)  =  \frac{1}{2\nu} - I_4(t, \phi),\nonumber
\end{eqnarray}
which is the symmetry independent equivalent of Eq. (\ref{key}) with $\alpha = 0$.  Eq. (\ref{gkey}) determines $g$ and $\phi_{\pm}$, self-consistently, as functions of $[\eta,\nu, \chi]$ at each value of $t$.

\textbf{Large Gap Limit and $t-J$ Model Relation:}  The general development in Appendix D shows the existence of a large gap solution in proximity to an interaction dependent singularity for any gap symmetry.  Using the same $Q \ll 1$ expansion of Eq. (\ref{gkey}), and the $t \rightarrow 0$ limits for the integrals given by Eq. (\ref{intex}), the $\phi_{+}(0)$ and $g(\phi)$ assume the asymptotic forms
\begin{equation}\label{qcp}
\phi^3   =  \gamma\nu\left(\frac{\nu/\chi - 2/3}{\eta - \nu}\right),\quad g =  1- \frac{\gamma}{\phi}.
\end{equation}
The gap expression is valid for $(2/3)\chi < \nu \leq \eta$.  It is evident that $\phi$ is singular at $\nu/\eta = 1$.  The unique point is attained for $\nu/\chi = U_0/\varepsilon_m > 2/3$, and $\nu/\eta = U_0/V_0 = 1$, independent of the density of states $N_0$.  Since $\chi \ll 1$ the point is reached even for extremely weak interactions -- a QCP characteristic.   Near the QCP, in the FOPT domain, both $\delta$ and $\Delta$ are very large, and $g \lessapprox 1$.  There is no order parameter that approaches zero near the QCP at the FOPT, as is the case for a QCP at a SOPT.\cite{varma}

In the QCP limit the scaled condensation energy $a\Delta\Omega(0, \phi) = a\Omega(0,\phi)$ saturates to the expression
\begin{equation}\label{Omegaqcp1}
a\Delta\Omega_o  \approx  1 - 2\left[1 + \frac{1}{2\chi}\right]\eta.
\end{equation}
The corresponding critical field $H_C(0)$ also saturates, and it follows from Eqs. (\ref{S}) and (\ref{Itlargephi}) that the entropy $S(t) \rightarrow 0$.  Although it has been conjectured that the observed spike in $\Delta\Omega$ near optimal doping\cite{tallon1} may signify proximity to a QCP, the comparisons in Section V indicate that cuprates are SOPT materials far from the large fluctuation regime that leads to a FOPT near the QCP.

In the large gap limit the the large gap Eq. (\ref{qcp}) is related algebraically to the three band $t-J$ model.  For $\nu \gg (2/3)\chi$, Eq. (\ref{qcp}) assumes the form
\begin{equation}\label{modeltj}
\Delta^3 (0)  = \frac{t^4}{J},\quad t  =  \sqrt{\varepsilon_m U_0},\quad N_0J = \frac{V_0 - U_0}{V_0 + U_0}.
\end{equation}
Setting $\Delta = E_p - E_d$, which is the cuprate charge transfer gap, Eq. (\ref{modeltj}) is identical to the analogous relation in the three band $t-J$ model for particle hopping between the Cu3d and the O2p orbitals.\cite{lee}  The relation between the models is remarkable, but difficult to interpret, particularly since the phonon interaction $V_0$ is not in the $t-J$ model.

\textbf{Large Fluctuation Numerical Solutions}

The solution $\phi_{-}(t)$ of Eq. (\ref{gkey}) coincides with the solution $\phi(t)$ of Eq. (\ref{keysm}) for small $p = \nu/\eta = U_0/V_0$.  For fixed $[\eta, \chi]$, increasing $p$ increases $\phi_{-}(t)$ until the SOPT solution $\phi_{-}(t_C) = 0 $ is lost and a FOPT solution $\phi_{+}$ emerges.  The FOPT occurs with $\phi_{+}(t_{SN}) \geq 0$ at the SN transition with maximum temperature $t_{SN}$ and gap $\phi_{+}(t_{SN}) > 0$.  In the FOPT domain there is also a $t_C \leq t_{SN}$ with $\phi_{+}(t_C) = 0$.  For $ t_C \leq t \leq t_{SN}$ there are two solutions for each value of $t$.  Hence, the FOPT exhibits $t$ dependent hysteresis with no applied magnetic field.  In the limit $p \rightarrow 1$, both $|\Omega(0, \phi)|$ and $t_{SN}$ saturate, but $\phi(t < t_{SN}) \rightarrow \infty$.  Since this unique point exists for all $t \leq t_{SN}$, and depends only on the ratio $p = U_0/V_0$, it satisfies the essential conditions of a QCP.\cite{tallon1}

\begin{figure}
\includegraphics[width = 3.2in]{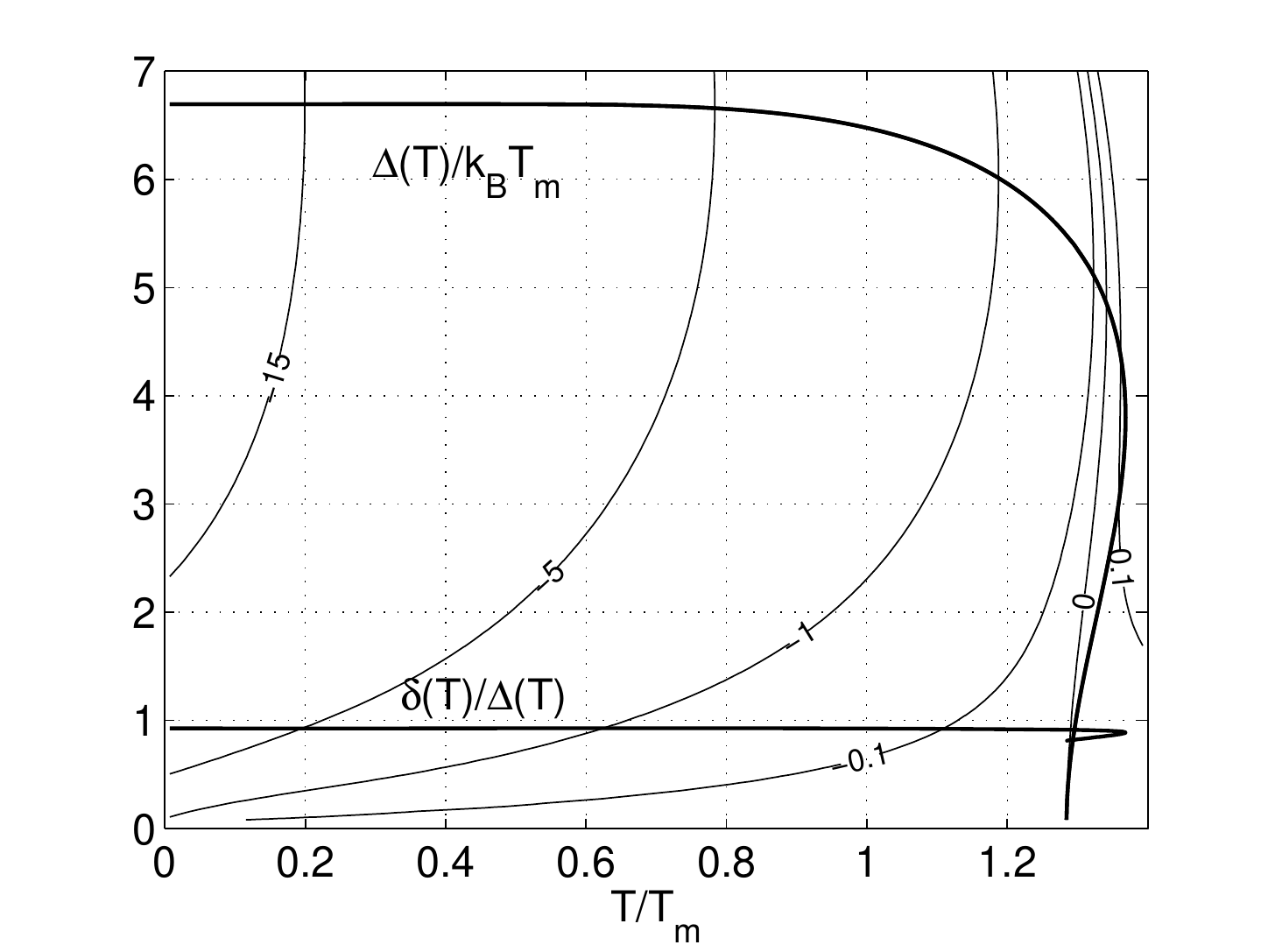}
\caption{\label{Fig10} FOPT solutions $\phi_{+}(t)$ and $|\delta(t)|/|\Delta(t)|$ of Eq. (\ref{gkey}) are plotted for  $\nu = 0.964\eta$, $\eta = 0.25$ and $\chi = 0.01$.}
\end{figure}

Fig. \ref{Fig10} shows a FOPT solution $\phi_{+}(t)$ of Eq. (\ref{gkey}).  Evident is the FOPT hysteresis, with the SN transition at $\phi_{+}(t_{SN}) \neq 0$ a higher $t$ than the NS transition at $\phi_{+}(t_C) = 0$.  The maximum $T$ saturates near $T \lessapprox 1$ as the QCP is approached.  This gap-$T_C$ decoupling contrasts the behavior in the SOPT domain, where both the gap and $T_C$ are increasing functions of $\nu$.  For very large $T_C$ the ratio $T_C/T_F \approx \chi \sim 10^{-1}-10^{-2}$, consistent with low carrier densities in HTS cuprates, and $10^3-10^4$ times the LTS $T_C/T_F$ values.\cite{tsuei}  The condensation energy contours near $t =0$ tend toward the QCP saturation value $a\Delta\Omega_o = - 24.5$ calculated from Eq. (\ref{Omegaqcp1}).

\begin{figure}
\includegraphics[width = 3.2in]{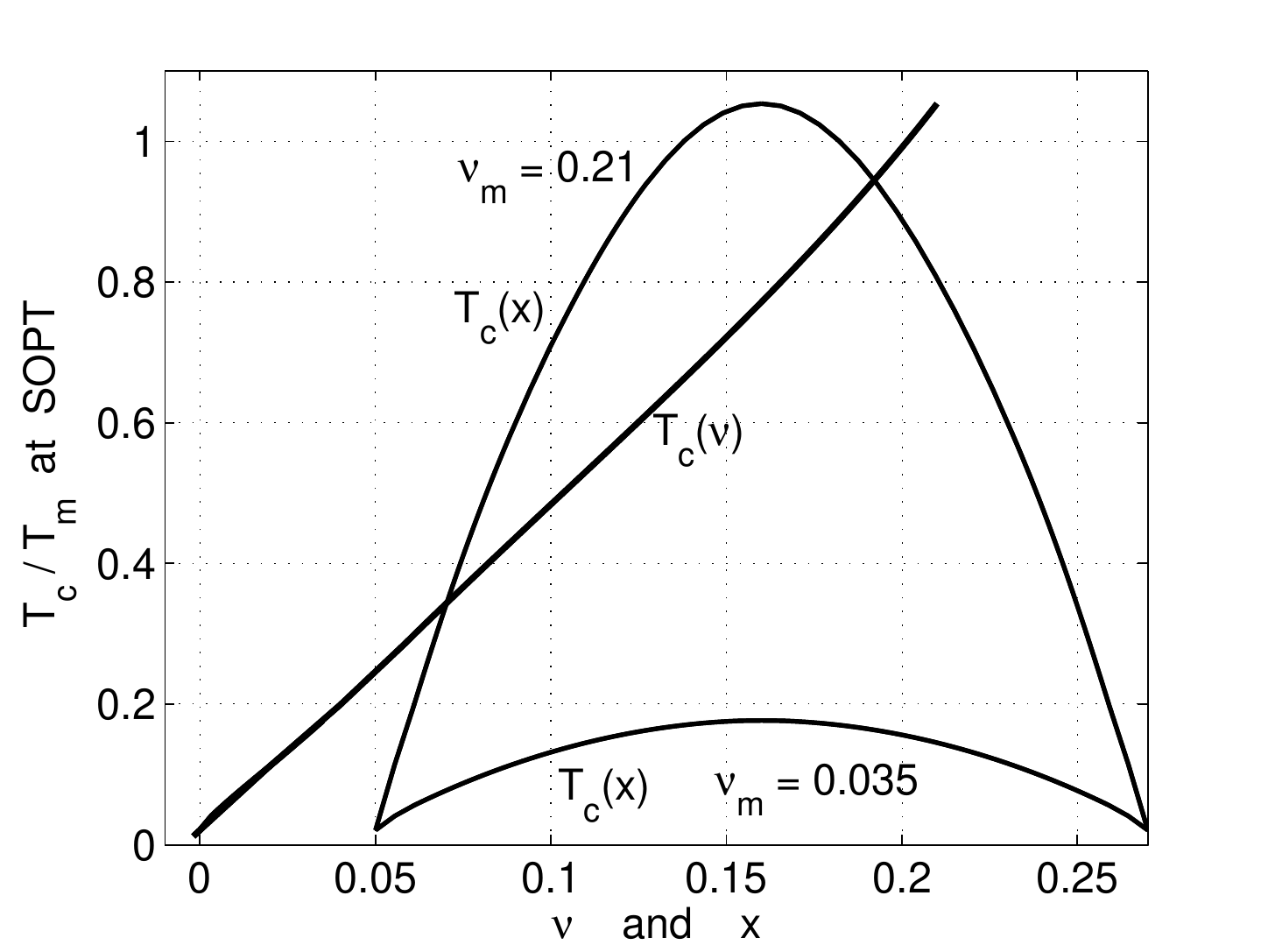}
\caption{\label{Fig11}   Plotted are SOPT phase boundaries $t_c(x)$ versus the hole doping concentration $x$ for $\eta = 0.25$, $\chi = 0.01$.  The upper boundary is for $\nu_m = 0.84\eta$ which is very close to the FOPT region, and the lower boundary is for $\nu_m = 0.14\eta$.}
\end{figure}

SOPT phase boundaries are plotted in Fig. \ref{Fig11}.  The upper $t_C(x)$, plotted for $\nu_m = 0.84\eta$, lies just below the FOPT domain.  The curve is slightly pinched near the maximum $T_C(x_{op})/T_m = 1.06$, where $T_C(\nu)$ is nonlinear.  At $x_{op}$ the enhancement factor is $48$ times the BCS MF value.  For comparison, the lower curve is plotted for the same parameters used in the typical cuprate phase boundary in Fig. \ref{Fig4}.  The value of $t_C(x_{op}) = 0.177$ is about $7\%$ higher than that in Fig. \ref{Fig4} where fluctuation was neglected.   A significant characteristic is the linearity of $T_C(\nu)$ over a wide range of $\nu$ values, which produces the parabolic phase boundary $T_C(x)$ observed in cuprates.

\begin{figure}
\includegraphics[width = 3.2in]{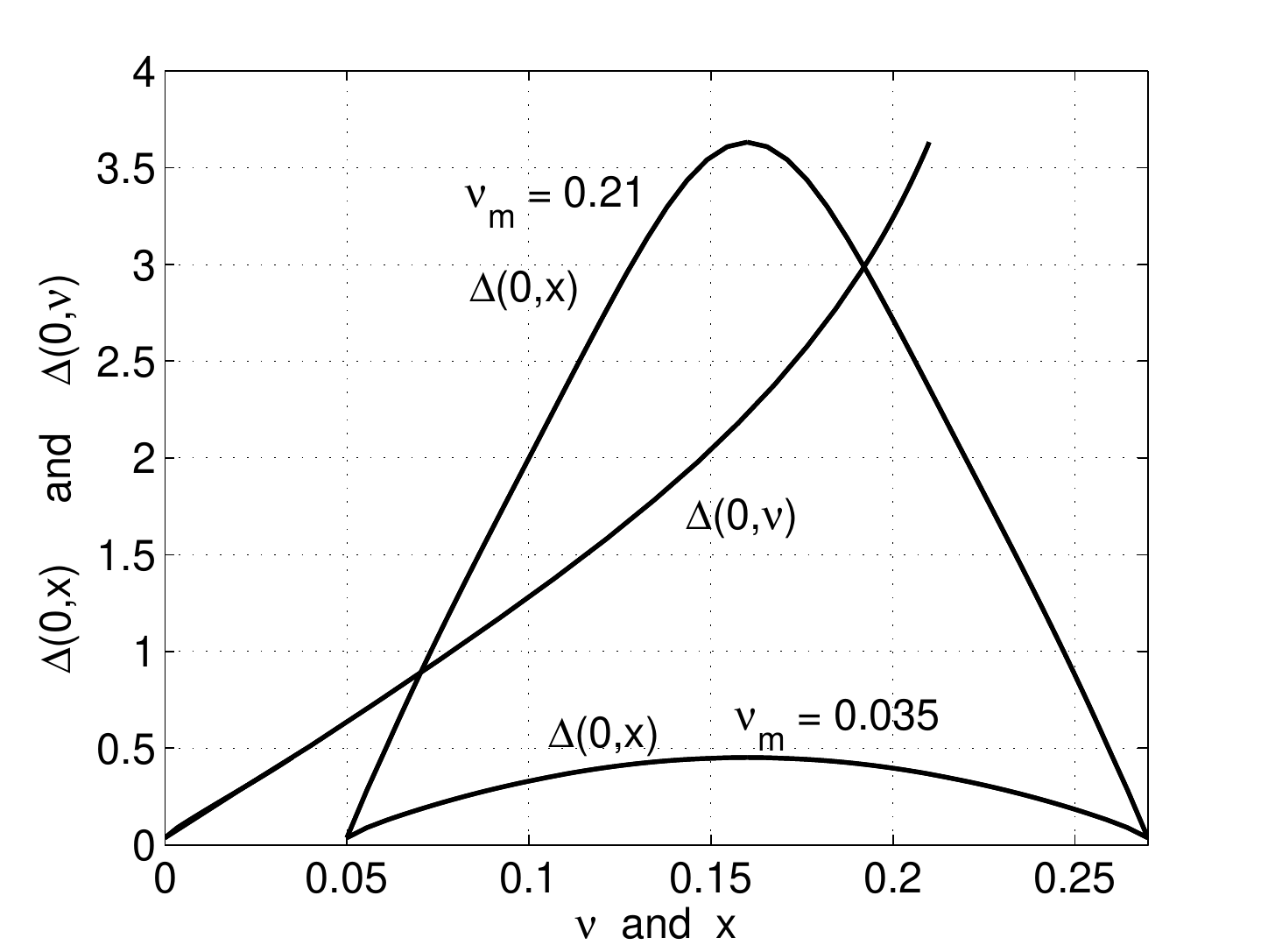}
\caption{\label{Fig12}   The SC gap $\Delta(T = 0, x)$ as a function of the hole doping concentration $x$, and as a function of the exchange parameter $\nu$ for same parameter values in Fig. \ref{Fig11}.}
\end{figure}
The SC gap $\Delta(T = 0)$ is plotted in Fig. \ref{Fig12} as $\Delta(\nu)$ and as $\Delta(x)$.  Linearity of $\Delta(\nu)$ over a relatively broad range of $\nu$ values, gives a parabolic $\Delta(x)$.  The upper curve $\Delta(x)$ is for parameter values very close to the FOPT domain.  As the QCP is approached, the increased nonlinearity of $\Delta(\nu)$ gives the triangular shape, which increasingly fills the phase domain below the pseudo- and ant-pseudo gaps plotted in Fig. \ref{Fig1}.  The lower curve is for same parameter values used in Fig. \ref{Fig5}, based on Eq. (\ref{keysm}) where fluctuation is neglected.   For the lower curve, $\Delta(x_{op}) = 0.451$ and $\Delta(x_{op})/k_BT_C = 2.55$.  The quantitative effect of the fluctuations is obtained by recalculating the value of $T_m$ from the experimental data, as done in connection with Fig. \ref{Fig5}.  This gives $T_m = 537$ from the experimental gap, and $T_m = 523$ from the experimental $T_C$.  The relative difference in $T_m$ is $2.6\%$, compared with $4.3\%$.  The error in the gap ratio is $2\%$ compared with $4\%$.   Thus for $\nu = 0.035$ neglect of the fluctuation term is justified.

\section{CONCLUSIONS}
The combined state probability-Hamiltonian model predicts a rather large set of doping dependent HTS properties listed in the introduction, and fundamental properties of the normal state.  The relative success is attributed to the following properties of the model:
\begin{itemize}

\item The state probability model, defines the probability of the SC and normal states based particle occupation of a large set of unit cell states.  To preserve local charge neutrality in the CuO plane, the SPM mandates the formation of distinct charge-spin plaquettes.  The SPM determines the doping dependence of the SC pairing interaction and the SC gap, and in the normal state it determines the pseudogap and an anti-pseudogap.

\item The procedure introduced to treat the deviation from the mean field Hamiltonian produces an exact expression for the internal energy ${\cal U}(T) = \langle H\rangle$.   The thermodynamic potential $\Omega(T)$ is exact at $T = 0$.

\item The phase average RPA value of the magnitude $|\delta|$ of static fluctuation from the mean field is not negligible for any interaction with non-zero diagonal matrix elements.  Values of $|\delta|$ and the SC gap $|\Delta|$ are determined self-consistently at the minimum free energy.

\item The spin-singlet exchange interaction $U$, with ${\bf k}$ independent diagonal matrix elements, is a very effective pairing glue, resulting in significant static fluctuation from the mean field state.

\item The SC gap and the critical temperature $T_C$ are linear functions of the exchange interaction parameter $\nu \propto U$ over a broad range of values of $\nu$, independent of the effective density of states and the kinetic energy cutoff parameter.

\item Large static fluctuation produces quantum criticality with concomitant extreme HTS properties, although relevance to cuprates is tentative.

\end{itemize}

The theory captures the key mechanism responsible for a large SC gap, high $T_C$, a large  $\Delta(T = 0)/k_BT_C$ ratio, and a low temperature $\lambda^{-2} \propto T$.  It also gives the doping dependent mechanism, coupled with $H$, that is responsible for the HTS phase transition boundary $T_C(x)$, and gap $\Delta(x, 0)$ for a broad range of cuprates exhibiting a SOPT with relatively high $T_C$.  The doping dependent probability model ascribes a physical basis to a "universal parabolic" function for $T_C(x)/T_C(x_{op}) = \Delta(0, x)/\Delta(0, x_{op})$ satisfied by cuprates.  It includes the empirical function\cite{presland, tallon2, hufner} as a special case.  The phonon interaction parameter $\eta \propto V$ gives only a mean field contribution the HTS state.  Universality stems entirely from the exchange parameter $\nu(x)$.   The form of $\nu(x)$ is further supported by the concomitant universal SC gap-pseudogap relation, which agrees with all observed parabolic phase boundaries.  Fitting the broad spectrum of cuprate data in Ref. \cite{hufner} further confirms the relevance and internal consistency of the theory.  These non-trivial experimental signatures substantiate the essential role of $U(x)$ as a pairing mechanism.   Symmetry of the state probability model suggests the possibility of HTS in electron doped "anti-cuprates", with the roles of the cation and anion reversed.

A FOPT with $T$ dependent hysteresis and large static fluctuation occurs in proximity to a QCP.  The QCP is independent of the gap symmetry function, the density of states, and the energy cut-off parameter.  The unique dependence on the interaction ratio $U_0/V_0$ with the emergence of quantum criticality, may indicate the possibility of extreme HTS with weak interactions.

Although the theory predicts many cuprate HTS properties, a number of aspects are not included in the Hamiltonian.  Doping dependence of parameters other than $\nu(x)$, and explicit contributions from the charge reservoir layers are neglected.  A fundamental evaluation of the state dependent contributions to the effective exchange $J(x)$, and the phonon mediated $V(x)$ is lacking.
\bibliographystyle{unsrt}
\bibliography{htsbib}

\appendix

\section{INTERACTION FORMULATION}
This appendix contains two parts.  The first relates to the exchange interaction, averaged over the UC states using the probability model in Section II. The second relates to the determination of the interaction matrix elements in the Hamiltonian in Section III.

\textbf{Average Exchange}.  Referring to Section II, we consider a checkerboard pattern of unit cell plaquettes$(\pm)$ with opposite net charge-spin.   The UC states in a given cell are denoted by $\varphi_{h\uparrow}, \varphi_{e\uparrow}$, and those in the neighbor cell are denoted by $\varphi_{h\downarrow}, \varphi_{e\downarrow}$, which satisfy the probability constraints in Eq. (\ref{Probplaqstates}) to maintain local charge-spin neutrality.  Formulation of the exchange interaction between states in the paired cells requires an extremely complicated microscopic determination of the overlap integrals for the $4 \times 16$ constituent UC states, which in turn determine the eigenstates of the interacting system.  Here only the relative values of the exchange for different groups of accessible states are formulated.  In the context of the density functional approach with a Kohn-Sham exchange correlation potential,\cite{callaway} the exchange values are assumed to be proportional to differences between characteristic energies $v_{h\sigma}(ij)$ and $-v_{e\sigma}(ij)$ corresponding to the UC states in Eq. (\ref{plaqstates}), as illustrated in Table \ref{hstates}.  In the SC state, excluding the UC  states $\varphi(11)$ and $\varphi(44)$, the average energies for the paired plaquettes are
\begin{eqnarray}
E_{h\uparrow} + E_{e\downarrow}& = & P_h^d P_e P_h [J_1P_eP_h + J_2P_e^2 + J_3P_h^2],\label{Ehup}\\&&\nonumber\\
E_{e\uparrow} + E_{h\downarrow} & = & -P_e^d P_e P_h[J_1^{\prime}P_eP_h + J_2^{\prime}P_h^2 + J_3^{\prime}P_e^2],\label{Eeup}\\&&\nonumber\\
J_k & = & V_k^{h\uparrow} - V_k^{e\downarrow}, \quad J_k^{\prime} = V_k^{e\uparrow} - V_k^{h\downarrow} \nonumber\\&&\nonumber\\
V_1^{q\sigma} & = & [v_{11} + v_{22} + 2(v_{12} + v_{34})]^{q\sigma}, \quad q = e, h\nonumber\\&&\nonumber\\
V_2^{q\sigma} & = & 2[v_{13} + v_{23}]^{q\sigma}, \quad V_3^{q\sigma} = 2[v_{14} + v_{24}]^{q\sigma}.\nonumber
\end{eqnarray}
The designation $e\sigma$ or $h\sigma$ in $V_k$ is implicit in every $v_{ij}$.  The $P^d$'s are Cu3d orbital occupation probabilities, and all $P$'s are O2 p-orbital occupation probabilities.

Adding  Eqs. (\ref{Ehup}) and (\ref{Eeup}), defining $J = E_{h\uparrow}+ E_{e\uparrow} + E_{h\downarrow} + E_{e\downarrow}$, and noting that $J_k = -J_k^{\prime}$ leads to the expression
\begin{eqnarray}\label{J}
\frac{J}{P_eP_h} & = & J_1P_eP_h + J_2[P_e^2 + P_h^2] -\nonumber\\&&\nonumber\\
 			&& (J_2 - J_3)[P_e^dP_e^2 + P_h^dP_h^2]
\end{eqnarray}\label{Jij}
Since $J_k = V_k^{h\uparrow} - V_k^{e\downarrow}$ each energy difference
\begin{equation}
J_{ij}  =  v_{h\uparrow}(ij) - v_{e\downarrow}(ij) = -[v_{e\uparrow}(ij) - v_{h\downarrow}(ij)].
\end{equation}
is interpreted as an effective spin-singlet exchange interaction between the states $\varphi_{h\uparrow}(ij)$ and $\varphi_{e\downarrow}(ij)$.  Noting that $J_2 \approx J_3$, and using $(P_e + P_h)^n = 1$, gives Eq. (\ref{Jx}).

\textbf{Matrix Elements of U and V:}  The matrix elements of $U \propto J$ as a spin-singlet exchange interaction is formulated.  A singlet exchange interaction between spins at coordinates $\bf r$ and $\bf {r}^{\prime}$ has the form
\begin{eqnarray}\label{Hexr}
H_{ex} & = & \frac{1}{4}\sum_{\bf r \bf {r}^{\prime}}2J_{{\bf r r}^{\prime}}\Xi^{\dagger}_{{\bf r r}^{\prime}}\Xi_{{\bf r r}^{\prime}}\nonumber\\& & \\
	& = & \sum_{\bf r \bf {r}^{\prime}}J_{{\bf r r}^{\prime}}[S_{\bf r} \cdot S_{{\bf r}'} - (1/4)n_{\bf r}n_{{\bf r}'}],\nonumber\\& & \nonumber\\
\Xi_{{\bf r r}^{\prime}} & = & c_{{\bf r}^{\prime}\downarrow} c_{{\bf r}\uparrow} - c_{{\bf r}^{\prime} \uparrow}c_{{\bf r}\downarrow}\nonumber
\end{eqnarray}
Setting ${\bf r}^{\prime} = {\bf r} + {\bf R}$, the operator $\Xi_{{\bf r r}^{\prime}}^{\dagger}$ creates a spin-singlet with a translationally invariant energy exchange constant $J_{{\bf r}, {\bf r} + \bf R} = J_{\bf R}$.  Eq. (\ref{Hexr}) and the transform of $H_{ex}$ to momentum space, with compaction to a single spin term is given without proof in Ref. \cite{paramekanti}.  We develop the transform details to point out an oversight in the determination of the diagonal elements, which are important for HTS.

Defining the Fourier transform
\begin{equation}\label{ftc}
c_{\bf {r} \sigma_i} = \frac{1}{\sqrt{N}}\sum_{\bf {k}_i}c_{\bf {k}_i \sigma_i}\exp{[i\bf {k_i}\cdot\bf r]},\quad i = 1,...,4
\end{equation}
and the shift ${\bf k}_1 = -{\bf k} + {\bf q}/2, {\bf k}_2 = {\bf k} + {\bf q}/2,
{\bf k}_3 = -{\bf k}^{\prime} + {\bf q}/2, {\bf k}_4 = {\bf k}^{\prime} + {\bf q}/2$, the transform of Eq. (\ref{Hexr}) is
\begin{eqnarray}\label{Hexk}
H_{ex} & = & \frac{1}{4N}\sum_{{\bf kk}^{\prime}}2J_{{\bf kk}^{\prime}}\Xi^{\dagger}_{\bf kq}\Xi_{\bf kq}\\& & \nonumber\\
\Xi_{\bf kq} & = & c_{-{\bf k} + {\bf q}/2\downarrow}c_{{\bf k} + {\bf q}/2\uparrow} - c_{-{\bf k} + {\bf q}/2\uparrow}c_{{\bf k} + {\bf q}/2\downarrow}\nonumber\\ & &\nonumber\\
    & = &  c_{-{\bf k} + {\bf q}/2\downarrow}c_{{\bf k} + {\bf q}/2\uparrow} + c_{{\bf k} + {\bf q}/2\downarrow}c_{-{\bf k} + {\bf q}/2\uparrow}\nonumber\\ & &\nonumber\\
J_{{\bf kk}^{\prime}} & = & \sum_{\bf R}J_{\bf R}\exp{[-i({\bf k} - {\bf k}^{\prime})\cdot\bf R]}\nonumber.
\end{eqnarray}
The $\bf k$ space representation of $H_{ex}$ is simplified by writing $J_{{\bf kk}^{\prime}}  =  J_{{\bf kk}^{\prime}}^e + J_{{\bf kk}^{\prime}}^o$, where
\begin{eqnarray}\label{Jsym}
J_{{\bf kk}^{\prime}}^e & = & J_{-{\bf k},{\bf k}^{\prime}}^e = J_{{\bf k}, - {\bf k}^{\prime}}^e,\nonumber\\ & &\\
J_{{\bf kk}^{\prime}}^o  & = & -J_{-{\bf k}, {\bf k}^{\prime}}^o = -J_{{\bf k}, -{\bf k}^{\prime}}^o.\nonumber
\end{eqnarray}
Noting the second form of $\Xi_{\bf kq}$ in Eq. (\ref{Hexk}), it follows that terms involving $J_{{\bf kk}^{\prime}}^o$ do not contribute to $H_{ex}$; whereas terms involving $J_{{\bf k}^{\prime}}^e$ combine to give a single spin set form
\begin{eqnarray}\label{Hexk1}
H_{ex} & = &  \frac{2}{N}\sum_{{\bf kk}^{\prime}\bf q}J_{{\bf kk}^{\prime}}^ep_{\bf k q}^{\dagger} p_{{\bf k}^{\prime}\bf q}\\& & \nonumber\\
p_{\bf k q} & = & c_{-{\bf k} + {\bf q}/2\downarrow}c_{{\bf k} + {\bf q}/2 \uparrow}.\nonumber
\end{eqnarray}

For a square lattice, applicable to the CuO plane of cuprates, we assume the nearest neighbor plaquette exchange $J_{\bf R} = -J$ is an average value for the $x$ and $y$ directions.  Setting  $R_x = R_y = 1$, the transformed exchange parameter is
\begin{eqnarray}\label{Jk}
J_{{\bf kk}^{\prime}} & = & -4J[C_{\bf{kk}'} +  S_{\bf{kk}'}],\\ & & \nonumber\\
C_{\bf{kk}'} & = & \cos{k_x}\cos{k_x^{\prime}} + \cos{k_y}\cos{k_y^{\prime}},\nonumber \\ & & \nonumber\\
S_{\bf{kk}'} & = & \sin{k_x}\sin{k_x^{\prime}} + \sin{k_y}\sin{k_y^{\prime}}\nonumber
\end{eqnarray}
Since $C_{\bf kk} +  S_{\bf kk} = 2$, the even part of $J_{{\bf kk}^{\prime}}$ is
\begin{equation}\label{Jk1}
J_{{\bf kk}^{\prime}}^e = -4J\left[(1-\delta_{\bf{kk}'})C_{\bf{kk}'} + 2\delta_{\bf{kk}'}\right].
\end{equation}
Defining $d$ and $s$-wave symmetry functions
\begin{equation}\label{dsdef}
\psi_{\bf k}^d =  \cos{k_x} - \cos{k_y}, \quad \psi_{\bf k}^s  =  \cos{k_x} + \cos{k_y},
\end{equation}
it follows that
\begin{equation}\label{ds}
2C_{\bf{kk}'} = \psi_{\bf k}^d\psi_{{\bf k}'}^d + \psi_{\bf k}^s\psi_{{\bf k}'}^s .
\end{equation}
This substitution of symmetry factors is used in Ref.\cite{paramekanti}, however we note that the diagonal elements are not correct. The contribution of $S_{\bf kk}$ to $J_{{\bf kk}^{\prime}}^e$ was overlooked.  This term, which produces the $\bf k$ independent diagonal element in $J_{{\bf kk}^{\prime}}$, is essential to HTS.

It is convenient to introduce a complex symmetry factor $\psi_{\bf k}$ defined as
\begin{equation}\label{psi}
\psi_{\bf k} = \frac{1}{\sqrt{2}}[\psi_{\bf k}^d + i \psi_{\bf k}^s],
\end{equation}
which gives the product
\begin{equation}\label{psi2}
\psi_{\bf k}\psi_{\bf {k}^{\prime}}^* = C_{\bf{kk}'} + i[\cos{k_y}\cos{k_x^{\prime}} -\cos{k_x} \cos{k_y^{\prime}}].
\end{equation}
It will be shown that all thermodynamic properties are functions only of $|\psi_{\bf k}|^2 = C_{\bf{kk}}$.   Furthermore, as stated in Ref. \cite{paramekanti}, the s-wave and d-wave contributions to $H_{int}$ are the same.

Using the complex $\psi_{\bf k}$, we replace the $J_{\bf k \bf {k}^{\prime}}^e$ with the complex exchange interaction
\begin{equation}\label{Jkc1}
U_{\bf k \bf {k}^{\prime}} = -U_0\left[(1-\delta_{\bf{kk}'})\psi_{\bf k}\psi_{\bf {k}^{\prime}}^* + 2\delta_{\bf{kk}'}\right],
\end{equation}
where $U_0 = 4J$.  As formulated in Section II, $J$ is replaced with an effective exchange $J(x)$ that depends on the doping dependent particle occupation of the Cu3d-O2p orbitals.

In the phonon mediated interaction $V$, the main contribution is from tight binding, small polarons formed in response to the Cu3d-O2p hopping of electrons(holes).  Drawing from the discussion above Eq. (\ref{Gamma}):  The symmetry of $V_{{\bf kk}^{\prime}}$ is the same as that of the exchange $J$, and the diagonal matrix elements $V_{\bf kk} = 0$.  The structure of the off-diagonal matrix elements, other than the symmetry factors, is a complicated function of $\bf k$ and ${\bf k}^{\prime}$, which is modeled here by a constant $-V_0$.  Accordingly, the effective contribution of $V$ to the electronic Hamiltonian is approximated by
\begin{equation}\label{Vkc1}
V_{\bf k \bf {k}^{\prime}} = -V_0(1-\delta_{\bf{kk}'})\psi_{\bf k}\psi_{\bf {k}^{\prime}}^*,
\end{equation}
giving a $V + U$ interaction
\begin{equation}\label{Gamma1}
\Gamma_{{\bf kk}^{\prime}} = -(V_0 + U_0)(1-\delta_{\bf{kk}'})\psi_{\bf k}\psi_{{\bf k}^{\prime}}^* - 2U_0\delta_{{\bf kk}'}.
\end{equation}

\section{DIAGONALIZATION OF H}

The Bogoliubov-Valatin canonical transformation,\cite{bogoliubov,valatin} to a new set of Fermion operators $\gamma_{\bf k }$ and $\lambda_{\bf k }$ is
\begin{eqnarray}\label{bvtransform}
\gamma_{\bf k } & = & u_kc_{\bf k } - v_{\bf k }c_{-\bf k }^\dagger, ~~	\lambda_{\bf k } =  u_{\bf k }c_{-\bf k } + v_{\bf k }c_{\bf k }^\dagger\nonumber\\& &\\			
c_{\bf k}& = & u_{\bf k}^{*}\gamma_{\bf k} + v_{\bf k }\lambda_{\bf k}^{\dagger},~~~  c_{-\bf k}^{\dagger}  =  -v_{\bf k}^{*}\gamma_{\bf k} + u_{\bf k}\lambda_{\bf k}^{\dagger},\nonumber
\end{eqnarray}
where the coefficients satisfy  $|u_{\bf k}|^2 + |v_{\bf k}|^2 = 1$.  The operators $\gamma_{\bf k }^\dagger(\gamma_{\bf k })$ and $\lambda_{\bf k}^{\dagger}(\lambda_{\bf k})$ create(destroy) quasi-particle excitations consisting of a correlated electron-hole pair.

Applying (\ref{bvtransform}) to $H_{kin}$ in Eq. (\ref{H1}), noting that $\varepsilon_{\bf k}  =  \varepsilon_{-\bf k}$ , and using the anti-commutation rules, gives
\begin{eqnarray}\label{H0tf}
H_{kin} & = & \sum_{\bf k } \varepsilon_{\bf k}[\hat{N}_{\bf k}^o + \nonumber \\
& & 2u_{\bf k}v_{\bf k}(\lambda_{\bf k}\gamma_{\bf k})^\dagger + 2(u_{\bf k}v_{\bf k})^*\lambda_{\bf k}\gamma_{\bf k}]\nonumber\\& &\\
\hat{N}_{\bf k}^o & = & (|u_{\bf k}|^2 - |v_{\bf k}|^2)(n_{\bf k} + q_{\bf k}) + 2|v_{\bf k}|^2,\nonumber	
\end{eqnarray}
where $n_{\bf k} = \gamma_{\bf k}^\dagger\gamma_{\bf k}$, and $q_{\bf k} = \lambda_{\bf k}^{\dagger}\lambda_{\bf k}$ are quasi-particle number operators.  To transform $H_b$ we use
\begin{eqnarray}\label{pairtf}
c_{-\bf k }c_{\bf k} = u_{\bf k}^*v_{\bf k}(1 - n_{\bf k} - q_{\bf k}) + \nonumber \\
(u_{\bf k}^*)^2\lambda_{\bf k}\gamma_{\bf k}  - (v_{\bf k})^2(\lambda_{\bf k}\gamma_{\bf k })^\dagger ,	
\end{eqnarray}
giving
\begin{eqnarray}\label{Hmftf}
H_b & = & \sum_{\bf k}\{ -(u_{\bf k}v_{\bf k}^*\Delta_{\bf k} + u_{\bf k}^*v_{\bf k}\Delta_{\bf k}^*)(1 - n_{\bf k} - q_{\bf k}) + \nonumber\\&{}\nonumber\\
 & & [(v_{\bf k}^*)^2\Delta_{\bf k} - (u_{\bf k}^*)^2\Delta_{\bf k}^*]\lambda_{\bf k}\gamma_{\bf k}  - \nonumber\\&{}\nonumber\\
& & [(u_{\bf k})^2\Delta_{\bf k} - (v_{\bf k})^2\Delta_{\bf k}^*](\lambda_{\bf k}\gamma_{\bf k})^\dagger + b_{\bf k}^*\Delta_{\bf k}\},  					
\end{eqnarray}
where $\Delta_{\bf k} = - \sum_{\bf k^\prime}\Gamma_{\bf k \bf {k}^\prime} b_{\bf k^\prime}$.
The off-diagonal terms $\lambda_{\bf k}\gamma_{\bf k}$ and $(\lambda_{\bf k}\gamma_{\bf k})^\dagger$ are eliminated from $H_0 = H_{kin} + H_b$ using the coefficient constraint
\begin{equation}\label{uvcon}
2\varepsilon_{\bf k}u_{\bf k}v_{\bf k}  - u_{\bf k}^2\Delta_{\bf k} + v_{\bf k}^2\Delta_{\bf k}^* = 0.					
\end{equation}

The solution of Eqs. (\ref{uvcon}) and $|u_{\bf k}|^2 + |v_{\bf k}|^2 = 1$ leads to the relations
\begin{eqnarray}\label{uv}
(u_{\bf k}/v_{\bf k})\Delta_{\bf k} & = & \varepsilon_{\bf k} \mp E_{\bf k},\quad
2E_{\bf k} u_{\bf k}^*v_{\bf k}  =  \mp \Delta_{\bf k},\nonumber\\& &\\
2E_{\bf k}|u_{\bf k}|^2 & = & E_{\bf k} \mp \varepsilon_{\bf k},\quad
2E_{\bf k}|v_{\bf k}|^2  =  E_{\bf k} \pm \varepsilon_{\bf k},\nonumber		 	
\end{eqnarray}
where $E_{\bf k} = \sqrt{\varepsilon_{\bf k}^2 + |\Delta_{\bf k}|^2}$.  Using Eqs. (\ref{uvcon}) and (\ref{uv}) in $H_{kin} + H_b(\Gamma)$ leads to the diagonal mean field form
\begin{equation}\label{Hmf}
H_{mf}  =  \sum_{\bf k}[\pm E_{\bf k }(1 - n_{\bf k} - q_{\bf k}) + \varepsilon_{\bf k} + \Delta_{\bf k}b_{\bf k}^*].
\end{equation}
It is stated in the literature that one should choose the lower sign in (\ref{uv}), but all thermodynamic functions are invariant with respect to the choice of sign, which is simply a choice of an electron or a hole picture.

\textbf{Determination of $\langle H_d\rangle$:} To complete the diagonalization of $H$ the operator $d_{\bf k}^\dagger d_{\bf k^\prime}$ is approximated by its average $\langle d_{\bf k}^\dagger d_{\bf k^\prime}\rangle$.  Define $X[\langle p_{\bf k} q_{\bf k^\prime}\rangle] = \langle p_{\bf k} q_{\bf k^\prime}\rangle - \langle p_{\bf k}\rangle\langle q_{\bf k^\prime}\rangle $.  Using the definition $d_{\bf k} = b_{\bf k} - c_{-\bf k}c_{\bf k}$ and applying Eq. (\ref{pairtf}), the $ b_{\bf k}$'s cancel and one obtains
\begin{eqnarray}\label{dd1}
X[\langle d_{\bf k}^\dagger d_{\bf k^\prime}\rangle] & = & X[\langle (c_{-\bf k}c_{\bf k})^\dagger c_{-\bf k^\prime}c_{\bf k^\prime}\rangle] \nonumber \\&{}\nonumber &\\
& = & u_{\bf k}v_{\bf k}^*u_{\bf k^\prime}^*v_{\bf k^\prime} X[\langle (n_{\bf k} + q_{\bf k})(n_{\bf k^\prime} + q_{\bf k^\prime})\rangle] + \nonumber\\&{}\nonumber &\\
& & (u_{\bf k}u_{\bf k^\prime}^*)^2\langle(\lambda_{\bf k}\gamma_{\bf k})^\dagger \lambda_{\bf k^\prime}\gamma_{\bf k^\prime}\rangle + \nonumber\\&{}\nonumber &\\
& & (v_{\bf k}^*v_{\bf k^\prime})^2\langle\lambda_{\bf k}\gamma_{\bf k}(\lambda_{\bf k^\prime}\gamma_{\bf k^\prime})^\dagger\rangle			
\end{eqnarray}
The average of all other terms in $X[\langle d_{\bf k}^\dagger d_{\bf k^\prime}\rangle ]$ involving unmatched creation and annihilation operators are zero in the eigenstates of $H_{mf}$.  Using the anti-commutation relation for Fermion operators to rearrange the last two terms in Eq. (\ref{dd1}), and noting that $\langle\gamma_{\bf k^\prime}^\dagger\gamma_{\bf k}\rangle = \delta_{\bf k k^\prime}\langle n_{\bf k}\rangle$ and $\langle\lambda_{\bf k^\prime}^\dagger\lambda_{\bf k}\rangle = \delta_{\bf k k^\prime}\langle q_{\bf k}\rangle$ gives

\begin{eqnarray}\label{dd2}
X[\langle d_{\bf k}^\dagger d_{\bf k^\prime}\rangle ] & = &  u_{\bf k}v_{\bf k}^*u_{\bf k^\prime}^*v_{\bf k^\prime} X[\langle (n_{\bf k} + q_{\bf k})(n_{\bf k^\prime} + q_{\bf k^\prime})\rangle] + \nonumber \\&{}\nonumber &\\
& & \delta_{\bf k k^\prime}|u_{\bf k}|^4\langle n_{\bf k}q_{\bf k}\rangle + \\&{}\nonumber &\\
& & \delta_{\bf k k^\prime}|v_{\bf k}|^4 \langle(1 - n_{\bf k})(1 - q_{\bf k})\rangle ,\nonumber
\end{eqnarray}
where $\delta_{\bf k \bf {k}^{\prime}} = 0$ for $\bf k \neq \bf {k}^{\prime}$ and $\delta_{\bf k \bf k} = 1$.  Since $n_{\bf k}$ and $q_{\bf k}$ are uncorrelated it follows from Eq. (\ref{average}) that $\langle n_{\bf k}q_{\bf k^\prime}\rangle = \langle n_{\bf k}\rangle\langle q_{\bf k^\prime}\rangle$ for all $k$ and $k^\prime$. Similarly, $\langle n_{\bf k}n_{\bf k^\prime}\rangle = \langle n_{\bf k}\rangle\langle n_{\bf k^\prime}\rangle$ for $k \neq k^\prime$, but $\langle n_{\bf k}n_{\bf k}\rangle  = \langle n_{\bf k}\rangle$, and $\langle q_{\bf k}q_{\bf k}\rangle  = \langle q_{\bf k}\rangle$, since the eigenvalues are 0 and 1.  Using these relations, one obtains $X[\langle (n_{\bf k} + q_{\bf k})(n_{\bf k^\prime} + q_{\bf k^\prime})\rangle] = \delta_{\bf k k^\prime}[\langle n_{\bf k}\rangle(1 - \langle n_{\bf k}\rangle + \langle q_{\bf k}\rangle(1 - \langle q_{\bf k}\rangle]$.  Finally, noting that $\langle n_{\bf k}\rangle = \langle q_{\bf k}\rangle$, Eq. (\ref{dd2}) reduces to
\begin{eqnarray}\label{dd}
X[\langle d_{\bf k}^\dagger d_{\bf k^\prime}\rangle ] & = & \frac{1}{4}\delta_{\bf k k^\prime}\langle \hat{N}_{\bf k}\rangle^2\\& & \nonumber\\
\frac{1}{2}\langle \hat{N}_{\bf k}\rangle  & = & |u_{\bf k}|^2\langle n_{\bf k}\rangle + |v_{\bf k}|^2(1 - \langle n_{\bf k}\rangle)\nonumber,
\end{eqnarray}
The $\langle N_{\bf k}\rangle = \langle N_{\bf k}^o\rangle$ is the average non-interacting fermion gas particle number density for state $\bf k$ for both spin orientations.  Using Eq. (\ref{uv}) in Eq. (\ref{dd}) and noting that $\langle d_{\bf k}^\dagger d_{\bf k^\prime}\rangle = \langle d_{\bf k}^\dagger\rangle\langle d_{\bf k^\prime}\rangle + X[\langle d_{\bf k}^{\dagger} d_{\bf k^\prime}\rangle ]$ gives $\langle H_d\rangle$ in Eq. (\ref{H}). In obtaining Eq. (\ref{H}) we apply Eq. (\ref{average}) to obtain $\langle n_{\bf k}\rangle = \langle q_{\bf k}\rangle = f(E_{\bf k}) = [e^{\beta E_{\bf k }} + 1]^{-1}$, and use $1 - 2f(x) = \tanh(x/2)$.
\section{THERMODYNAMIC FUNCTIONS}
General relations between several thermodynamic functions are derived from their basic definitions, and different definitions of specific heats are related.

Grand canonical ensemble average of an operator $Q$:
\begin{eqnarray}\label{average}
\langle Q \rangle & = & Tr(\hat{\rho} Q),~~~\hat{\rho} = \frac{1}{Z}\exp (-\beta H),\nonumber\\& &\\
Z & = & Tr[\exp (-\beta H)],\nonumber
 \end{eqnarray}
where $\beta = 1/(k_B T)$, $H(\mu) = H(0)-\mu \hat{N}$ with chemical potential $\mu(T)$, $Z$ is the grand partition function, and $\hat{\rho}$ is the density operator.

Thermodynamic potential (generalized free energy):
\begin{equation}\label{defOmega}
\Omega = -(1/\beta)\ln(Z).
\end{equation}

Von Neumann entropy:
\begin{equation}\label{defentropy}
S = -k_BTr(\hat{\rho}\ln\hat{\rho)}.
\end{equation}
The above definitions (\ref{average}) - (\ref{defentropy}) give the entropy
\begin{equation}\label{entropy}
S = \frac{1}{T}(\langle H\rangle - \Omega) = -\frac{\partial\Omega}{\partial T} + \langle \frac{\partial H}{\partial T}\rangle.
\end{equation}

The partition function for $H$ in Eq. (\ref{H}) is
\begin{eqnarray}\label{Z}
Z & = & \prod_{\bf k}[1 + \exp(-\beta E_{\bf k })]^2 \exp[\beta(E_{\bf k } - \varepsilon_{\bf k})]\times\nonumber\\
& &\exp(-\beta \Delta_{\bf k}b_{\bf k}^*)\exp(-\beta \langle H_d\rangle).
 \end{eqnarray}
Using $Z$ one obtains the expressions for $\Omega$ in Eq. (\ref{Omegak}), and the internal energy ${\cal U} = \langle H\rangle$ in Eq. (\ref{Uk}).

\textbf{Specific Heat:} Differentiating S in Eq. (\ref{entropy}) with respect to $T$ and eliminating $S$, yields the useful derivative relation
\begin{equation}\label{dtSH}
T\frac{\partial S}{\partial T} = \frac{\partial \langle H\rangle}{\partial T}- \langle\frac{\partial \hat{H}}{\partial T}\rangle = -T\frac{\partial^2\Omega}{\partial T^2} + T\frac{\partial}{\partial T} \langle\frac{\partial H}{\partial T}\rangle.
\end{equation}
The various forms of specific heat in Eq. (\ref{dtSH}) are
\begin{eqnarray}
C & = & T\frac{\partial S}{\partial T},\label{C}\\
  &  & \nonumber\\
C_{\cal U} & = & \frac{\partial \langle H\rangle}{\partial T} = C + \langle\frac{\partial H}{\partial T}\rangle,\label{CU}\\
  & & \nonumber\\
C_{\Omega} & = & -T\frac{\partial^2\Omega}{\partial T^2} = C - T\frac{\partial}{\partial T} \langle\frac{\partial H}{\partial T}\rangle.\label{COmega}
\end{eqnarray}
Details for $C$ are given in Eq. (\ref{C1}).  It is evident that for any model with $\langle\partial H/\partial T\rangle \neq 0$, the specific heats differ.  In this case the internal energy $\cal U$ cannot be determined from an integration of $C$ with respect to $T$, as done for LTS.\cite{tinkham}

\textbf{Condensation Energy at Low Temperature:} The condensation energy is $\Delta\Omega(t, \phi) = \Omega(t, \phi) - \Omega(t, 0) \leq 0$, where $\Omega(t, \phi)$ is given by Eq. (\ref{Omega}) In the limit $2t \ll \phi^2$, and $2t \ll 1$ for $\phi = 0$, the integrals in Appendix E give $\Omega (t, \phi) \approx \Omega (0, \phi)$ to within exponentially small $t$ dependence, and
\begin{equation}\label{Omegat0}
a\Omega(t, 0) = -\frac{2\nu}{\chi}(\ln 4 - 1)t - \frac{\pi^2}{3}t^2.
\end{equation}
Noting that $\Delta\Omega(0, \phi) = \Omega(0, \phi)$, leads to the expression
\begin{eqnarray}\label{condens2}
\frac{\Delta\Omega (t, \phi)}{\Delta\Omega (0, \phi)}& = & 1 - 2B(\phi)\frac{T}{T_C} - A(\phi)\left(\frac{T}{T_C}\right)^2,\\
& &\nonumber\\
B(\phi) & = & \frac{\nu}{\chi}\frac{\ln 4 - 1}{|a\Omega(0, \phi)|}\frac{T_C}{T_m}\nonumber\\
 & &\nonumber\\
A(\phi) & = & \frac{\pi^2}{3|a\Omega(0, \phi)|}\left(\frac{T_C}{T_m}\right)^2.\nonumber
\end{eqnarray}
The factor $\nu/\chi$ in Eq. (\ref{condens2}) shows that the linear $T$ dependence is due to the diagonal matrix elements of $U$.  For $\nu = 0$ the LTS dependence $(T/T_C)^2$ is recovered with $A \approx (2/3)(\pi T_C/\Delta)^2 \approx 2.12$.  But, even for a small value of $\nu$ the linear term in $T/T_C$  dominates, since $\chi \ll 1$.  Eq. (\ref{condens2}) gives critical field Eq. (\ref{Hc}).

\section{GENERAL ANALYSIS}
The purpose of this appendix is to outline the behavior of the general, symmetry dependent minimum free energy solutions.  The $\Omega$ corresponding to the model $H$ in Eq. (\ref{Hmod}) is
\begin{eqnarray}\label{Omegasd}
\Omega  & = & \Omega_{mf} + \langle H_d\rangle,\\& & \nonumber\\
\Omega_{mf} & = & - \frac{4}{\beta}\Sigma_0 + \sum_{\bf k}\varepsilon_{\bf k} + \frac{|\Delta |^2}{\Gamma_0},\nonumber\\ & &\nonumber\\
\langle H_d\rangle & = & - \Theta_0\frac{|\delta|^2}{\Gamma_0^2} - U_0\Sigma_1,\quad \Theta_0  =  2(1 + \alpha) U_0, \nonumber
\end{eqnarray}
where $\alpha \ll 1$ is defined in Eq. (\ref{Sigmas}).  The constraint Eq. (\ref{Ddconk1}) is
\begin{equation}\label{Ddcon}
\pm\frac{|\delta|}{|\Delta|} = g = 1 - \Gamma_0\Sigma,\quad \Gamma_0 = V_0 + U_0,
\end{equation}
and the $\bf k$-space sums are
\begin{eqnarray}\label{sums}
\Sigma_0 & = & \frac{1}{2}\sum_{\bf k}\ln [2\cosh (\beta E_{\bf k} /2)],\nonumber\\&{}\nonumber &\\
\Sigma  & = & \frac{1}{2}\sum_{\bf k}\frac{|\psi_{\bf k}|^2}{E_{\bf {k}}}\tanh(\beta E_{\bf k }/2),\\& & \nonumber\\
\Sigma_1 & = & \frac{1}{2}\sum_{\bf {k}}\sigma_{\bf k}^2,\quad \sigma_{\bf k} = 1 - \frac{\varepsilon_k}{E_{\bf k }}\tanh(\beta E_{\bf k }/2),\nonumber
\end{eqnarray}
where $E_{\bf k} = \sqrt{\varepsilon_{\bf k}^2 + |\Delta_{\bf k}|^2}$ and $\Delta_{\bf k} = \Delta\psi_{\bf k}$.  It follows from Eqs. (\ref{Omegasd})-(\ref{sums}) that
\begin{eqnarray}\label{DOmegasd}
a\frac{\partial\Omega}{\partial |\Delta|^2}|_T & = & \frac{g}{\Gamma_0} - \Theta_0(1+\rho) \frac{g^2}{\Gamma_0^2} - \Theta_0\Sigma_4\frac{g}{\Gamma_0} - U_0\Sigma_2\nonumber\\& & \\
\rho & = & \frac{\alpha}{1 + \alpha}\frac{\partial\ln\alpha}{\partial\ln|\Delta|^2}.\nonumber
\end{eqnarray}
The sums arising from the derivatives of $\Sigma_1$ and $\Sigma$, respectively, are
\begin{eqnarray}
\Sigma_2 & = & \frac{1}{2}\sum_{\bf {k}}|\psi_{\bf k}|^2\frac{\varepsilon_k}{E_{\bf k }^2}\sigma_{\bf k}\zeta_{\bf k},\nonumber \\ & &\nonumber\\
\Sigma_4 & = & \frac{1}{2}\sum_{\bf {k}}|\psi_{\bf k}|^2\frac{|\Delta_{\bf k}|^2}{E_{\bf k }^2}\zeta_{\bf k},\\ & &\nonumber\\
\zeta_{\bf k} & = & \frac{1}{E_{\bf {k}}}\tanh(\beta E_{\bf k }/2)- \frac{\beta}{2}\cosh^{-2}(\beta E_{\bf k }/2).\nonumber
\end{eqnarray}
The first term in Eq. (\ref{DOmegasd}) is from the mean field $\Omega_{mf}$, and the remaining terms are from $\langle H_d\rangle$.  Replacing the symmetry factor $|\psi_{\bf k}|^2$ in $\Sigma_{d1}$, defined in Eq. (\ref{Sigmas}), by its average value unity, $\alpha = 0$ and $\rho = 0$.  Since $\alpha \approx 0$, it is treated as a parameter and $\rho \ll 1$ is neglected.

Solving $\partial\Omega/\partial |\Delta|^2 = 0$ for $g/\Gamma_0$ gives
\begin{eqnarray}\label{key}
\frac{1}{\Gamma_0} - \Sigma & = & \frac{W}{2\Theta_0}\left[1 \pm \sqrt{1 - Q}\right],\\& & \nonumber\\
Q & = & \frac{4U_0\Theta_0}{W^2}\Sigma_2,\quad W = 1 - \Theta_0\Sigma_4.\nonumber
\end{eqnarray}
Solutions of Eq. (\ref{key}) are the gap amplitudes $|\Delta_{\pm}(T)|$.   Eq. (\ref{key}) is complicated and it has several distinct solution domains depending on the relative values of the parameters.  Extreme solutions of Eq. (\ref{key}) are the small and large gap solutions that occur in the same limit $Q \ll 1$.  To linear order in $Q$, Eq. (\ref{key}) leads to
\begin{eqnarray}
\frac{\Gamma_0 - \Theta_0}{\Gamma_0\Theta_0} & = & -(\Sigma - \Sigma_4) + \frac{U_0}{W}\Sigma_2 \label{sQ+} \\ & & \nonumber\\
\Sigma & = & \frac{1}{\Gamma_0} - \frac{U_0}{W}\Sigma_2,\label{sQ-}
\end{eqnarray}
Eq. (\ref{sQ+}) determines $|\Delta_{+}|$ and Eq. (\ref{sQ-}) determines $|\Delta_{-}|$, corresponding to the sign in Eq. (\ref{key}).  Setting $U_0 = 0$,  Eq. (\ref{sQ-}) reduces to the BCS constraint for symmetry $\psi_{\bf k}$, with a small, LTS gap $|\Delta_{-}|$.  Since $\Sigma - \Sigma_4 \geq 0$, Eq. (\ref{sQ+}) has no finite real solution in the limit $U_0 \rightarrow 0$.

Eq. (\ref{sQ-}), with $W \approx 1$, is the small $U_0$ equation that follows from the MF part $\Omega_{mf}$ plus the diagonal interaction term proportional to $\Sigma_1$ in Eq. (\ref{DOmegasd}), neglecting terms generated by $|\delta|^2$.  As $U_0$ is increased from zero Eq. (\ref{sQ-}) has an effective $\Gamma_{eff} = \Gamma_0/(1 - U_0\Gamma_0\Sigma_2 > \Gamma_0$ that causes the gap amplitude $|\Delta_{-}|$ to increase exponentially from the BCS value. [See Eq. (\ref{expsol}).]  The term small $U_0$ is quantified by the condition
\begin{equation}\label{smallU}
\Theta_0[U_0\Sigma_2 + \Sigma_4] \ll 1,
\end{equation}
required for negligible fluctuation effect from $|\delta|^2$.

Retaining the fluctuation terms leads to the emergence of a second solution $|\Delta_{+}|$ determined from Eq. (\ref{sQ+}), which becomes large when $\Gamma_0 - \Theta_0 \gtrapprox 0$.  In this limit the large gap solution of Eq. (\ref{sQ+}) is
\begin{equation}\label{qcps}
|\Delta_{+}(0)|^3  = \frac{\Gamma_0\Theta_0}{V_0 - (1 + 2\alpha)U_0}\Sigma_M,
\end{equation}
\[\Sigma_M  =  \frac{1}{2}\sum_{\bf {k}}\frac{|\psi_{\bf k}|^2\varepsilon_k}{(\varepsilon_{\bf k}^2/|\Delta_{+}|^2 + \psi_{\bf k}^2)^{3/2}}\left[U_0\sigma_{\bf k} - \varepsilon_{\bf k}\right].\]
Existence of $|\Delta_{+}|$ requires a minimum value of $U_0$ such that $\Sigma_M > 0$.  Eq. (\ref{qcps}) is singular when the interaction ratio $p = U_0/V_0$ assumes the value $p_0 = 1/(1 + 2\alpha) \lessapprox 1$.  Near the singularity the gap $|\Delta_{+}(0)| \rightarrow \infty$, and the $\Sigma_M$ is essentially independent of $|\Delta_{+}|$.

Although $|\Delta_{+}|$ becomes increasingly large for $p$ near $p_0$, the $g$ saturates to its maximum $g = 1$, and $\Omega$ remains finite.  It follows from Eq. (\ref{Omegasd}), with some manipulation, that the asymptotic saturation value is
\begin{equation}\label{omegasat}
\Omega_o(0)  =  \sum_{{\bf k}}(\varepsilon_{\bf k} - U_0/2) - \frac{\Gamma_0}{4}\left(\sum_{\bf k}|\psi_{\bf k}|\right)^2,
\end{equation}
for any symmetry $|\psi_{\bf k}|$.

Several inferences are drawn from the analysis above:  The symmetry factor $\psi_{\bf k}$ does not fundamentally change the thermodynamic properties.  For small values of $U_0$, defined by (\ref{smallU}), the $|\Delta_{-}|$ is independent of $|\delta|$.  The model Hamiltonian with $|\delta|^2$ neglected is applied in Section III, and its relevance to cuprates is clearly manifested by extensive comparison with experiment in Section V.   When the inequality (\ref{smallU}) is violated, retention of $|\delta|^2$ produces a gap solution governed by $p = U_0/V_0$ with a QCP at $p_0 \lessapprox 1$.  Extreme HTS properties of the model are considered in Section VI.

\section{INTEGRALS}
The integrals in Eqs. (\ref{Omega}) - (\ref{gcon}), and (\ref{gkey}), with integration variable $y = \langle\varepsilon_{\bf k}\rangle/\varepsilon_m$, are
\begin{eqnarray}\label{integrals}
I_0 (t, \phi)& = & \int_{0}^{1}dy \ln \left[2\cosh\left(\frac{Y}{2t}\right)\right],\quad Y = \sqrt{y^2 + \phi^2}\nonumber\\&{}\nonumber &\\
I(t, \phi) & = & 2\frac{t}{\phi}\frac{\partial I_0}{\partial\phi}\nonumber =
      \int_{0}^{1}dy\frac{1}{Y}\tanh \left(\frac{Y}{2t}\right),\nonumber\\&{}\nonumber &\\
I_1 (t, \phi) & = & \frac{1}{2}\int_{0}^{1}dy\left[1-\frac{y}{Y}\tanh\left(\frac{Y}{2t}\right)\right]^2,\nonumber\\& &\\
I_2(t, \phi)& = & \frac{1}{\phi}\frac{\partial I_1}{\partial\phi} =   \int_{0}^{1}dy\frac{y}{Y^2}\left[1-\frac{y}{Y}\tanh\left(\frac{Y}{2t}\right)\right]\times \nonumber\\&{}\nonumber &\\
& & \left[\frac{1}{Y}\tanh\left(\frac{Y}{2t}\right) - \frac{1}{2t}\cosh^{-2}\left(\frac{Y}{2t}\right)\right]\nonumber\\& &\nonumber\\
I_3(t, \phi) & = & \int_{0}^{1}dyY\tanh \left(\frac{Y}{2t}\right)\nonumber\\&{}\nonumber &\\
I_4(t, \phi) & = & -\phi\frac{\partial I}{\partial\phi} =  \phi^2\int_{0}^{1}dy\frac{1}{Y^2}\left[\frac{1}{Y}\tanh\left(\frac{Y}{2t}\right)\right.  -\nonumber\\&{}\nonumber &\\
& & \left.\frac{1}{2t}\cosh^{-2}\left(\frac{Y}{2t}\right)\right].\nonumber
\end{eqnarray}
Integration of $I_2$ by parts gives the useful form
\begin{eqnarray}\label{I22}
2I_2(t, \phi) & = & \left[1-\frac{1}{Y_1}\tanh\left(\frac{Y_1}{2t}\right)\right]^2 - \nonumber\\& &\nonumber\\
 & & \left[1-\frac{2}{\phi}\tanh\left(\frac{\phi}{2t}\right)\right] - \nonumber\\& &\nonumber\\
& & \int_{0}^{1}dy\left[\frac{1}{Y}\tanh\left(\frac{Y}{2t}\right)\right]^2,
\end{eqnarray}
The integrals satisfy the relations
\begin{eqnarray}\label{intrelate}
2t\frac{\partial I_0}{\partial t} + \frac{1}{t}I_3 & = & \frac{1}{2}I \frac{\partial \phi^2}{\partial t}, \nonumber\\& &\\
I\frac{\partial \phi^2}{\partial t} - 2\frac{\partial I_3}{\partial t} & = & \frac{1}{t^2}\int_{0}^{1}dy\left[Y^2 -\frac{t}{2}\frac{\partial \phi^2}{\partial t}\right]\cosh^{-2}\left(\frac{Y}{2t}\right).\nonumber\\ \nonumber
\end{eqnarray}

For $t \ll t_C$ the integrals are given by their $t = 0$ limit
\begin{eqnarray}\label{Itzero}
4tI_0(0, \phi) & = & Y_1 + \phi^2I(0, \phi),\quad Y_1 = \sqrt{1 + \phi^2},\nonumber\\&{}& \nonumber \\
I(0, \phi) & = & \mbox{arcsinh}(1/\phi) = \ln[(1/\phi)(1+ Y_1)]\nonumber\\&{}& \nonumber \\
I_4(0, \phi)& = & 1/Y_1,\\&{}&\nonumber  \\
I_1(0, \phi) & = & 1- Y_1 + \phi[1 - (1/2)\arctan(1/\phi)]\nonumber\\&{}&\nonumber \\
I_2(0, \phi)& = & -(1/Y_1)[1 - 1/(2Y_1)] + \nonumber\\&{}&\nonumber\\
 & & (1/\phi)[1 - (1/2)\arctan(1/\phi)].\nonumber	
\end{eqnarray}
For $\phi \gg 2t$,
\begin{equation}\label{Itlargephi}
2tI_0(t, \phi) = I_3(t, \phi) = \int_{0}^{1}dyY
\end{equation}
Expansions of the integrals in Eq. (\ref{Itzero}) in powers of $1/\phi \ll 1$ are
\begin{eqnarray}\label{intex}
I(0, \phi) & = & \frac{1}{\phi}-\frac{1}{6\phi^3},~~~I_4(0, \phi) = \frac{1}{\phi} -\frac{1}{2\phi^3} ,\nonumber\\&{}&\\
I_1(0,\phi) & = & \frac{1}{2} - \frac{1}{2\phi} +\frac{1}{6\phi^2}, ~~~I_2(0, \phi) =  \frac{1}{2\phi^3}.\nonumber
\end{eqnarray}
In the limit $\phi(t) \rightarrow 0$
\begin{eqnarray}\label{Iphizero}
I_1(t, 0) & = & 1 - t[\tanh{(1/2t)} + 2\ln \cosh{(1/2t)}],\nonumber\\& &\\
I_4(t, 0)& = & 0.\nonumber
\end{eqnarray}
For $t \lessapprox 0.1$, with $\gamma_e = 0.57726$, giving $D = 2.2677$,
\begin{eqnarray}\label{Iphi0}
I(t, 0) & = & \int_{0}^{1/2t}\frac{dx}{x}\tanh(x) \approx  \ln{(D/2t)},\nonumber\\& & \\
\ln{D} &\approx & -\int_{0}^{\infty}dx\ln{x}\cosh^{-2}{x} = \ln{(4/\pi)} + \gamma_e ,\nonumber\\& &\nonumber\\
I_2(t, 0)	& \approx & \frac{C_1(t)}{t},\quad C_1(t) = \frac{1}{2} - \frac{1}{4}\int_0^{1/2t}\frac{dx}{x^2}\tanh^2(x). \nonumber
\end{eqnarray}

\end{document}